\documentclass[reprint,prl,aps,superscriptaddress]{revtex4-2}
\usepackage{times}
\usepackage{graphicx}
\usepackage{amsmath, braket, amsfonts}
\usepackage{amssymb}
\usepackage{natbib}
\usepackage{sidecap}
\usepackage{bm, color, ulem}
\usepackage{xcolor}
\usepackage{ragged2e}
\usepackage{wrapfig,lipsum,booktabs}
\usepackage{siunitx}

\newcommand{\be}{\begin{equation}}
\newcommand{\ee}{\end{equation}}

\DeclareUnicodeCharacter{03BD}{{$\nu$}}
\DeclareUnicodeCharacter{2212}{{-}}

\makeatletter
\def\maketitle{
\@author@finish
\title@column\titleblock@produce
\suppressfloats[t]}
\makeatother

\begin{document}
\title{Universal chiral Luttinger liquid behaviour in a graphene fractional quantum Hall point contact  }

\author{Liam A. Cohen}
\thanks{These authors contributed equally to this work}
\affiliation{Department of Physics, University of California at Santa Barbara, Santa Barbara CA 93106, USA}
\author{Noah L. Samuelson}
\thanks{These authors contributed equally to this work}
\affiliation{Department of Physics, University of California at Santa Barbara, Santa Barbara CA 93106, USA}
\author{Taige Wang}
\affiliation{Department of Physics, University of California, Berkeley, California 94720, USA}
\affiliation{Material Science Division, Lawrence Berkeley National Laboratory, Berkeley, California 94720, USA}
\author{Takashi Taniguchi}
\affiliation{International Center for Materials Nanoarchitectonics,
National Institute for Materials Science,  1-1 Namiki, Tsukuba 305-0044, Japan}
\author{Kenji Watanabe}
\affiliation{Research Center for Functional Materials,
National Institute for Materials Science, 1-1 Namiki, Tsukuba 305-0044, Japan}
\author{Michael P. Zaletel}
\affiliation{Department of Physics, University of California, Berkeley, California 94720, USA}
\affiliation{Material Science Division, Lawrence Berkeley National Laboratory, Berkeley, California 94720, USA}
\author{Andrea F. Young}
\email{andrea@physics.ucsb.edu}
\affiliation{Department of Physics, University of California at Santa Barbara, Santa Barbara CA 93106, USA}
\date{\today}
 \begin{abstract}
One dimensional conductors are described by Luttinger liquid theory, which predicts a power-law suppression of the density of states near the Fermi level. The scaling exponent is non-universal in the general case, but is predicted to be quantized for the chiral edge states of the fractional quantum Hall effect.  
Here, we report conductance measurements across a point contact linking integer and fractional quantum Hall edge states. At weak coupling, we observe the predicted universal quadratic scaling with temperature and voltage. At strong coupling, the conductance saturates to $e^2/2h$, arising from perfect Andreev reflection of fractionalized quasiparticles at the point contact.  We use the strong coupling physics to realize a nearly dissipationless DC voltage step-up transformer, whose gain of 3/2 arises directly from topological fractionalization of electrical charge.  
\end{abstract}

\maketitle

The Landau theory of Fermi liquids provides a near-ubiquitous description of interacting fermion systems.
One exception is provided when electrons are confined to one dimension, where arbitrarily weak interactions favor a distinct phase known as the Tomonaga-Luttinger Liquid\cite{luttinger_exactly_1963, tomonaga_remarks_1950, chang_chiral_2003, giamarchi_transport_2004}. 
In this phase, the low-energy collective excitations are orthogonal to the single-electron operators from which they are microscopically constructed.
This `orthogonality catastrophe' manifests experimentally as a power-law suppression of the electron tunneling density of states, $N(E) \propto (E-E_\text{F})^{1/g-1}$, at the Fermi energy $E_{\text{F}}$, despite the fact that the system remains conductive.  
The power law is characterized by an exponent $g$ known as the Luttinger parameter, which depends continuously on the nature and strength of the interparticle interactions\cite{fisher_transport_1997}.  
Experimentally, Luttinger liquid behavior can manifest through a non-Ohmic current-voltage relation $I(V) \propto V^{1/g}$, as observed, for example, in ropes of single-walled carbon nanotubes\cite{bockrath_luttinger-liquid_1999, ishii_direct_2003}.

An alternative means of creating a one-dimensional wire is at the boundary of a topologically ordered phase, as occurs in the fractional quantum Hall effect\cite{halperin_quantized_1982}.  
Here,  the right- and left- moving modes are physically separated to opposite edges of the two dimensional sample, resulting in a chiral  Luttinger liquid in which backscattering is suppressed entirely and the Luttinger parameter $g$ becomes quantized\cite{wen_chiral_1990}.   
In this setting, $g$ becomes a  fingerprint of the  topological order of the  enclosed bulk; for example $g = \frac{1}{3}$ for the Laughlin state at Landau level filling $\nu = \frac{1}{3}$\cite{wen_chiral_1990, kane_edge-state_1996}. 
As a result, tunneling of whole electrons into the fractionalized edge of the $\nu = \frac{1}{3}$ state is predicted to exhibit a quadratic scaling $G \propto T^{2}, V^{2}$. A central prediction of the theory is that this behavior is universal and independent of microscopic details at sufficiently low temperatures and bias voltages\cite{wen_chiral_1990}.

Experimentally, the most successful test of the chiral Luttinger liquid theory was obtained by studying tunneling between the edge of a two dimensional electron gas (2DEG) and a three dimensional electrode grown on the cleaved edge of the semiconductor wafer\cite{chang_observation_1996,chang_chiral_2003}. While striking power-law behavior was observed over a wide range of bias voltages and temperatures\cite{chang_observation_1996}, the exponent $1/g \sim 2.7$ was found to vary across samples and within the $\nu=1/3$ plateau\cite{grayson_continuum_1998, chang_plateau_2001}---in disagreement with the predicted quantized exponent $1/g=3$. 

\begin{figure*}
    \includegraphics[width = 179mm]{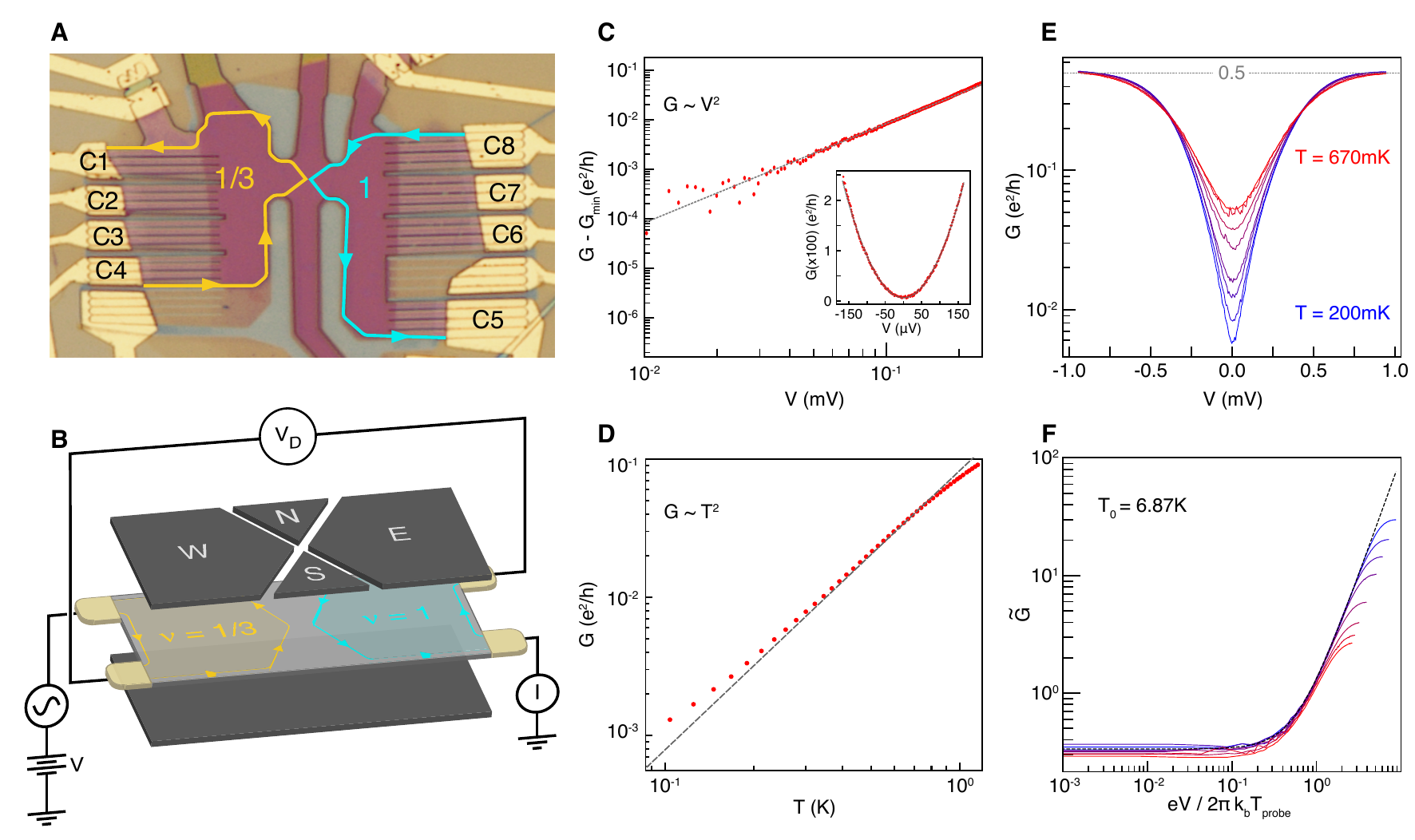}
    \caption{
\textbf{Universal conductance scaling at weak coupling.}
\textbf{(a)} Optical micrograph of the device with schematic depiction of chiral edge states. 
\textbf{(b)} Device schematic showing the patterned top graphite layer, graphene monolayer, and global bottom graphite gate\cite{cohen_tunable_2022}. 
The tunneling conductance across the junction is determined from the transmitted current $I$ and diagonal voltage $V_D$ as $G \equiv I / V_D$. 
\textbf{(c)} 
$G$ measured as function of $V$ at $T_{\text{probe}} = \SI{56}{mK}$, with $V_{\text{NS}} = −2.465V$ and $B = 10$~T. 
The inset shows a parabolic fit to the low-$V$ regime, giving $T_0=9.02\pm 0.007K$ as defined in Eq.~\eqref{eq:lutt_cond}.  
The main panel shows $G - G_{\text{min}}$, where $G_{\text{min}}=7.5\times 10^{-4} e^2/h$ is the minimum conductance. Fitting a power-law gives an exponent of $2.00 \pm 0.06$ \cite{cohen_supplementary_2022}, where the error represents the standard deviation in the fit parameter. 
\textbf{(d)} $G$ measured at $V=0$ as a function of temperature at the same gate voltages as panel (c).  
The dashed line is a plot of the conductance given by the first term of  Eq.~\eqref{eq:lutt_cond}, using $T_0=9.02$~K. 
\textbf{(e)} Nonlinear differential conductance for $T_{\text{probe}}$ =202mK, 245mK, 290mK, 344mK, 450mK, 549mK, 618mK, 666mK at $V_{\text{NS}} = −2.456V$.  
\textbf{(f)} 
The same data as in panel (e) after scaling $G$ and $V$ as described in the main text.  
The black dashed line is $\widetilde{G}$ as predicted by  Eq.~\eqref{eq:lutt_cond}.
    \label{fig:fig1}}
    \vspace{-10pt}
\end{figure*}

In principle, a quantum point contact between integer and fractional quantum Hall edge states\cite{wen_chiral_1990, lee_edge_1991, fisher_transport_1997, kane_edge-state_1996,chamon_distinct_1997, sandler_andreev_1998} provides a richer test-bed for chiral Luttinger liquid physics. 
In this geometry, the collective modes of the $\nu = \frac{1}{3}$ and $\nu=1$ edges may be modeled as chiral bosonic fields $\phi_{a}$ and $\phi_b$, respectively, 
coupled by a \textit{single} point scatterer of strength  $\Gamma$.  The low energy physics is described by the Lagrangian
\begin{equation}
\label{eq:lagrangian}
   \mathcal{L} = \frac{1}{4\pi} \sum_{i=a,b} \partial_x \phi_i (\partial_t - \partial_x) \phi_i      
 + \Gamma\delta(x)\left(\psi_a^\dagger\psi_b+\psi_b^\dagger \psi_a\right) 
\end{equation}
where the operators $\psi_a = e^{i \sqrt{3} \phi_a}$ and $\psi_b = e^{i \phi_b}$ remove an electron on the $\nu = 1/3$ and $\nu=1$ edges, respectively.  
In Eq.~\eqref{eq:lagrangian}, the first term describes the gapless bosonic edge modes on either side of the junction, while the second term describes inter-edge tunneling of electrons at the point contact.  
In the language of the renormalization group, the scaling dimensions of the electron operator $[\psi_b]=1/2$ while $[\psi_a]=3/2$  is three times larger, reflecting the topological order of the $\nu = 1/3$ fractional quantum Hall bulk.  In order for the corresponding 2D Euclidean action to remain dimensionless, $[\Gamma] = 1- [\psi_a] - [\psi_b] = -1$, meaning that edge-to-edge tunneling is irrelevant and becomes increasingly less important at low energies. This leads to the remarkable conclusion that no matter how ``open'' the junction is made---in other words, no matter how large the bare value of $\Gamma$ is--- the conductance will vanish at a sufficiently low temperature and voltage bias\cite{kane_transmission_1992}. 
Near this decoupled fixed point, the conductance can be computed within perturbation theory \cite{chang_chiral_2003}, giving
\begin{equation}
    \label{eq:lutt_cond}
    G (V, T) = \frac{e^2}{2h} (\frac{2 \pi T}{T_0})^2 \bigg [ \frac{1}{3} + (\frac{eV}{2 \pi k_b T})^2 + \cdots \bigg]
\end{equation}
for small temperatures $T$ and voltage bias $V$.
Here the bare tunneling strength is represented by $T_0$, where for weakly coupled edges $T_0 \propto 1/\Gamma$.
In this limit, the power law exponent $1/g - 1 = 2$ describing the $T$ and $V$ dependence provides a direct probe of the bulk topological order.  

Quantum point contact experiments in semiconductor quantum wells have indicated power-law behavior over a limited temperature range\cite{milliken_indications_1996,maasilta_line-shape_1997}; however, the presence of disorder at the tunnel junction complicates the physical interpretation \cite{milliken_indications_1996, maasilta_line-shape_1997,radu_quasi-particle_2008, miller_fractional_2007, hashisaka_andreev_2021}.  For both quantum point contact and cleaved-edge overgrowth experiments, the literature indicates that non-universal effects arising from the detailed structure of experimentally realized edges\cite{chamon_sharp_1994, wan_edge_2003, chamon_distinct_1997, kun_field_2003} push the universal scaling regime to energy scales beyond experimental reach.

Recently, graphene has emerged as a compelling platform for precision tests of chiral Luttinger liquid physics.  Most importantly, the low disorder available in field-effect controlled devices make a wide range of fractional quantum Hall states accessible\cite{zibrov_tunable_2017, dean_fractional_2020}.  
However, prior studies of quantum point contacts have been limited to regimes dominated by resonant tunneling effects, due at least in part to disorder introduced during the fabrication of local split gates\cite{zimmermann_tunable_2017,overweg_electrostatically_2018,deprez_tunable_2021,ronen_aharonov-bohm_2021}.  
Here, we use anodic oxidation lithography to pattern nanoscale features in graphite gates, which are then incorporated into a van der Waals  heterostructure to produce a clean quantum point contact \cite{cohen_tunable_2022}. 
A micrograph and schematic of our device is shown in Fig.~\ref{fig:fig1}A-B.  
The device architecture features a four-quadrant split gate geometry\cite{cohen_tunable_2022} where independent voltages may be applied to the North, South, East and West top gates ($V_{\text{N}}$, $V_{\text{S}}$, $V_{\text{E}}$ and $V_{\text{W}}$, respectively) and a global bottom gate ($V_{\text{BG}}$).  With $V_{\text{BG}}$ held constant, $V_{\text{E}}$ and $V_{\text{W}}$ fix the filling factors of the East and West regions to 1 and 1/3, respectively, while $V_{\text{NS}}=V_{\text{N}}=V_{\text{S}}$ is used to create a constriction by tuning the filling of the North and South regions to $\nu\leq0$.  
The choice of $V_{\text{BG}}$ controls the `sharpness' of the potential profile at the constriction, parameterized by the energy scale $E_V \equiv e \frac{\partial V}{\partial x} \ell_b$.  $E_V$ plays a key role in the physics of fractional quantum Hall edges: when the confinement energy $E_V$ is smaller than the Coulomb energy, the edge may reconstruct\cite{chamon_sharp_1994, wan_reconstruction_2002}, introducing additional edge modes which may push the universal tunneling behavior to experimentally inaccessible energy scales. As described in Refs. \onlinecite{cohen_tunable_2022,cohen_supplementary_2022}, the control available in our geometry allows us to access the universal regime of $E_C<E_V$ while maintaining independent control of the bulk filling factor and quantum point contact transparency.  

We begin by investigating the `weak-coupling' regime where $G \ll e^2/h$. 
This corresponds to the limit of $eV \ll k_b T_0$ and $T \ll T_0 / 2 \pi $ represented by Eq.~\eqref{eq:lutt_cond}.  
We measure $G \equiv I/V_D$ (see Fig.~\ref{fig:fig1}B), which yields the tunneling conductance while removing contributions from contact resistances\cite{cohen_tunable_2022, cohen_supplementary_2022}. 
We tune $T_0$ via $V_\mathrm{NS}$.
Fig \ref{fig:fig1}C shows $G$  measured as a function of the DC voltage bias, $V$, at a fixed $T_\mathrm{probe} = 56$mK and $V_\mathrm{NS} = -2.465V$. 
As seen in the inset, the $V$-dependence is well fit by a parabola, with a curvature corresponding to $T_0 = 9.02K$ in Eq.~\eqref{eq:lutt_cond}.  
To assess the quality of the power law fit, we subtract the minimum conductance, $G_{\text{min}}$ and plot $G-G_{\text{min}}$ on a logarithmic scale (Fig.~\ref{fig:fig1}C).  We find a simple $V^2$ power law over one order of magnitude in $V$, corresponding to two orders of magnitude in $G-G_{\text{min}}$. Specifically, we find an exponent of $2.00\pm.06$ \cite{cohen_supplementary_2022}. 
While we have not studied the filling factor dependence in detail, a second measurement taken at a different magnetic field and bulk filling shows the same exponent\cite{cohen_supplementary_2022}.

We next compare experiment with the predicted zero-bias temperature dependence, $G(V=0, T, T_0)$, of  Eq.~\eqref{eq:lutt_cond}.  The result for $T_0=9.02K$ is shown in Fig.~\ref{fig:fig1}D.
In contrast to the voltage dependence, a $T^2$ power law is  observed only for T between 200mK and 700mK.  We attribute deviations at lower temperatures to a decoupling of the electronic temperature from $T_\text{probe}$.  At high temperatures, deviations are expected as corrections to Eq.~\eqref{eq:lutt_cond}  become important, with significant deviations onsetting for $T > T_0/(4\pi)\approx \SI{750}{mK}$\cite{cohen_supplementary_2022}. The identical power laws for $V$ and $T$ are a manifestation of a general scaling relation.  Defining $\tilde G = 2 G \cdot \left(\frac{T_0}{2\pi T}\right)^2 \cdot \frac{h}{e^2}$ and $x=eV/\left(2\pi k_B T\right)$, it follows that $\tilde G=1/3+x^2 + \cdots$ provides a universal low-energy collapse.
Fig.~\ref{fig:fig1}E shows $G$ as a function of $V$ for several different temperatures at $V_\mathrm{NS} = -2.456V$. The same data, plotted as $\tilde G(x)$, is shown in Fig.~\ref{fig:fig1}F.  In these data sets, $T_0=6.87K$ is determined by fitting the lowest temperature curve in Fig.~\ref{fig:fig1}E to Eq.~\eqref{eq:lutt_cond}.  For low values of $V$ where universality is expected \cite{cohen_supplementary_2022}, the curves collapse onto the universal parabolic curve expected from chiral Luttinger liquid theory.

\begin{figure}[ht]
    \includegraphics[width = 90mm]{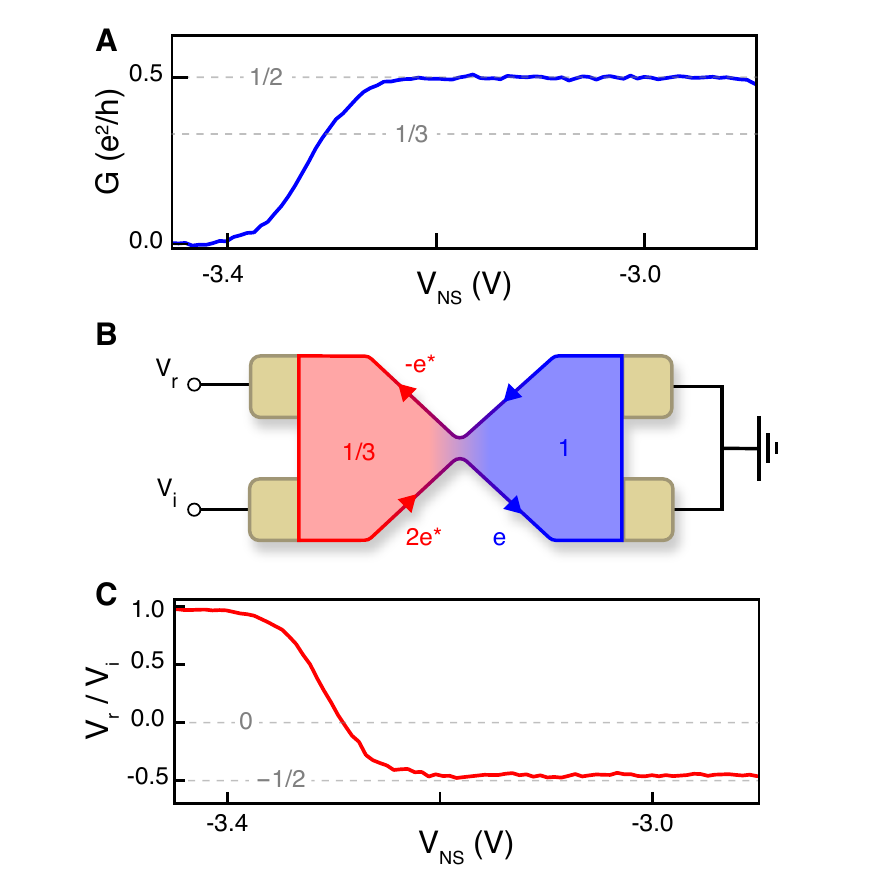}
    \caption{
    \textbf{Andreev-like quasiparticle scattering.} 
    \textbf{(a)} $G$ versus $V_\mathrm{NS}$ taken at $B= 9$T with a finite DC voltage bias of $145 \mu V$.  Here, $V_\mathrm{BG}$=2.0V,  $V_\mathrm{E}$=-1.460V, and $V_\mathrm{W}$=-1.775V to maintain $\nu = 1$ and $\nu = 1/3$ on the `East' and `West' sides of the junction respectively.  
    \textbf{(b)} Schematic representation of the Andreev scattering process for  fractionally charged quasiparticles in the strong coupling limit\cite{sandler_andreev_1998} of a $\nu =1/3$ to $\nu = 1$ point heterojunction.  
    \textbf{(c)} Ratio of the reflected voltage $V_\mathrm{r}$ to the source voltage $V_\mathrm{i}$ versus $V_\mathrm{NS}$; all other gate voltages are the same as in panel (a). 
    }
    \label{fig:andreev_reflection}
\end{figure}

As seen in Fig.~\ref{fig:fig1}E, $G$ approaches $\frac{e^2}{2h}$ at high bias. 
Naively, this is surprising: one might expect that full transmission of the incoming fractional edge mode would cause the conductance to saturate at $G = \frac{e^2}{3 h}$.  In fact, the observed $G\approx \frac{e^2}{2h}$ can be understood from the peculiar properties of the point contact at strong coupling, defined as $eV \gg k_b T_0$ or $T \gg T_0$.  The strong coupling regime is accessed most readily by lowering $T_0$ (i.e., increasing $V_{\text{NS}}$) at fixed $V$ and $T$, leading to a plateau in the differential $G\approx \frac{e^2}{2h}$  (Fig.~\ref{fig:andreev_reflection}A).  

Microscopically, the excess conductance can be understood by analogy to Andreev scattering at a metal-superconductor interface \cite{sandler_andreev_1998, chamon_distinct_1997, milovanovic_edge_1996}. 
In this picture, conduction occurs when a pair of incident quasiparticles, each with charge $e^* = -e/3$, is transmuted into a single electron on the $\nu=1$ side through the simultaneous retro-reflection of a charge-$e/3$ quasi-\textit{hole} into the downstream chiral edge state (see Fig.~\ref{fig:andreev_reflection}B).  This process leads to the observed nearly-quantized increase in $G$.  Moreover, when the Andreev process is the dominant form of charge transfer across the junction, the downstream $\nu=1/3$ edge hosting the retro-reflected hole is expected to develop a negative chemical potential with magnitude that is one half of the voltage of the incoming edge\cite{chklovskii_consequences_1998, sandler_andreev_1998}.  
Fig.~\ref{fig:andreev_reflection}C shows the measured reflected voltage $V_\mathrm{r}$ under the same conditions as the $G$ data in Fig.\ref{fig:andreev_reflection}A.  The near-quantization of both $G$ and $V_r/V_i$ over the same broad range of junction transparency implies that the described Andreev process completely dominates charge transfer.  
This contrasts with previous experiments in semiconductor wells\cite{hashisaka_andreev_2021}, where a much smaller effect was observed, likely mediated by resonant scattering.

\begin{figure*}[ht]
    \includegraphics[width = 180mm]{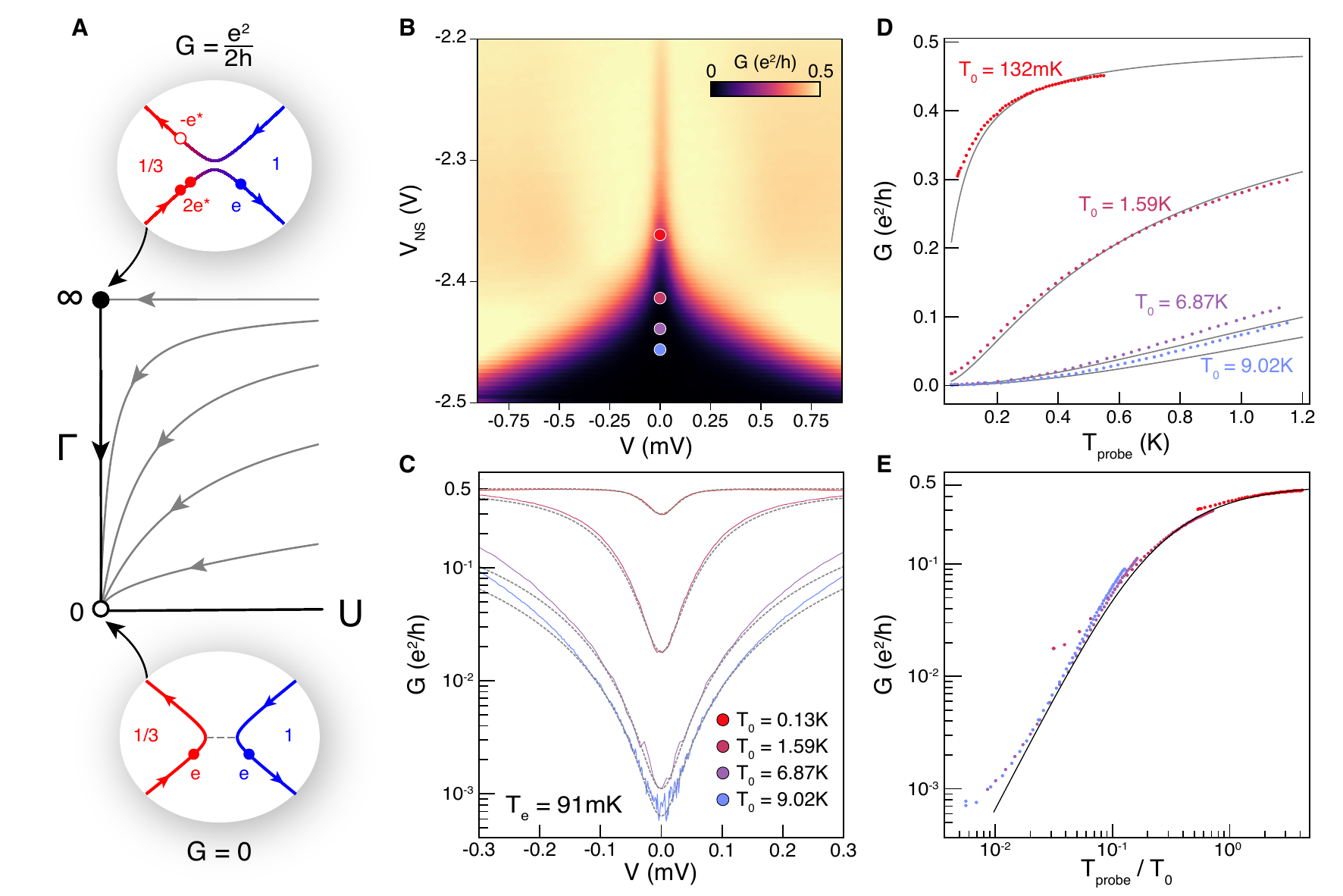}
    \caption{
    \textbf{Crossover from weak to strong coupling.}
    \textbf{(a)} Schematic of renormalization group flow in the $\Gamma$-$U$, where $U$ represents additional perturbations to Eq.~\eqref{eq:lagrangian}.  
    \textbf{(b)} $G$ as a function of the voltage on the North/South gates, $V_\mathrm{NS}$, and the DC voltage bias $V$ at $B=10T$.
    \textbf{(c)} Line cuts of panel (b) at the values of $V_{\text{NS}}$ indicated by the colored points.  Black dashed lines are plotted using the value of $G$ predicted by  Eq.~\eqref{eq:integral_conductance} where the parameter $T_0$ is extracted from the low-bias conductance.
    \textbf{(d)} The zero-bias conductance also scales with temperature in agreement with  Eq.~\eqref{eq:integral_conductance} for low energies. 
    While the data deviates from the model at high energies, $G$ nevertheless exceeds $G = \frac{e^2}{3h}$ for $T \gg T_0$, indicating strong coupling at high $T$.
    \textbf{(e)} The data from panel d, after scaling $T$ by $T_0$.  The curves collapse onto the universal scaling formula  Eq.~\eqref{eq:integral_conductance}, shown in black.
    \label{fig:strong_coupling}
    }
\end{figure*}

In our discussion, we have considered only single electron tunneling between edges, parameterized by $\Gamma$; however, additional processes such as electron co-tunneling may also contribute, which would be represented by additional operators not included in Eq.~\eqref{eq:lagrangian}.  
Renormalization group analyses have shown\cite{kane_edge-state_1996} that as  $\Gamma\rightarrow0$---these terms are more irrelevant than $\Gamma$. This guarantees that regardless of microscopic details Eq.~\eqref{eq:lagrangian} becomes a good approximation to the physical system at sufficiently low energies.  This is characteristic of a stable fixed point of the renormalization group, and accounts for the universal scaling behavior demonstrated in Fig.~\ref{fig:fig1}.

A different result is obtained at strong coupling---formally $\Gamma \rightarrow \infty$---which represents an unstable fixed point\cite{sandler_andreev_1998, fendley_exact_1994}. 
At this fixed point the dominant process that transfers charge across the junction is Andreev scattering. However, while other processes vanish when $\Gamma=\infty$---giving  $G=\frac{e^2}{2h}$---for any finite value of $\Gamma$, $G$ is expected to vanish at low energies as the system flows toward the stable $\Gamma=0$ fixed point. This is shown schematically in Fig.~\ref{fig:strong_coupling}A, which depicts a renormalization group flow diagram indicating the trajectory of the conductance as the energy is lowered.  In this plot, the y-axis, $\Gamma$, represents the coefficient of the operator which transfers a \textit{single} electron between the two edge modes, while the x-axis represents the coefficient `U' of any operator not explicitly captured in Eq.~\eqref{eq:lagrangian}. As is seen in the diagram, finite $U$, expected for realistic heterojunctions would seem incompatible with approaching the strong coupling fixed point.  For this reason, the  strong coupling limit was previously thought to be practically inaccessible except through highly tuned resonant scattering \cite{kane_resonant_1998,chklovskii_consequences_1998}.

The fact we observe near-perfect Andreev reflection when the junction is highly transmissive suggests this system approaches the strong-coupling fixed point with negligible contributions from the additional operators represented by $U$. 
When $U = 0$, Eq.~\eqref{eq:lagrangian} can be mapped to an integrable `quantum impurity model' with an exact solution for all values of $\Gamma$\cite{kane_transmission_1992,fendley_exact_1994}.  
The solution provides an expression for $G$ at arbitrary $V$, $T$, and $T_0$, 
\begin{equation}
    \label{eq:integral_conductance}
    \begin{split}
    &G(V, T) = \frac{e^2}{2 h} \int_{-\infty}^{\infty} dE \frac{-(2E+eV)^2}{(2E+eV)^2 + (k_b T_0)^2} f'(E)
    \end{split}
\end{equation}
where $f(E)$ is the Fermi-Dirac function\cite{fendley_exact_1994}. 
Eq.~\eqref{eq:integral_conductance} provides a universal crossover function which describes the transition between weak and strong coupling, depicted as a single line along the y-axis ($U = 0$) of Fig.~\ref{fig:strong_coupling}A\cite{chamon_distinct_1997, sandler_andreev_1998, fendley_exact_1994}. The integral reduces to Eq.~\eqref{eq:lutt_cond} for $T_0\rightarrow \infty$, corresponding to the weak coupling fixed point, and gives $\frac{e^2}{2h}$ for $T_0 \to 0$ corresponding to strong coupling. 

Fig \ref{fig:strong_coupling}B shows $G$ as a function of $V$ and $V_{\text{NS}}$ at a probe temperature $T_{\text{probe}} = \SI{56}{mK}$.  
Throughout the plotted range, the high $V$ conductance saturates to approximately $\frac{e^2}{2h}$ as expected for the strong coupling fixed point.  Meanwhile a zero-bias dip remains visible for all $V_{\text{NS}}$, indicative of the instability of the point contact to edge decoupling at low energies. 
To quantify how closely the quantum impurity model describes our system, we compare Eq.~\eqref{eq:integral_conductance} to our experimental data as a function of $V$ and $T$, and $T_0$. 
In Fig.~\ref{fig:strong_coupling}C, we plot four curves extracted for different values of $V_{\text{NS}}$.  For each curve, $T_0$ is determined from a fit to the low bias behavior with an appropriate low-energy expansion of Eq.~\eqref{eq:integral_conductance} \cite{cohen_supplementary_2022}. 
For the largest fit values of $T_0$, the residual value of $G$ when $V=0$ can then be used as a primary thermometer---in effect allowing us to correct for a possible lack of equilibration between the electron temperature and the probe thermometer. 
Using this method, we find $T_{\text{electron}} = \SI{91}{mK}$ for the $T_0=9.02K$ trace, in contrast to the measured probe temperature of 56mK. 
Taking  $\SI{91}{mK}$ as the electron temperature for the remaining data sets in Fig.~\ref{fig:strong_coupling}C, Eq.~\eqref{eq:integral_conductance} may then be used to generate $V$-dependent curves interpolating between weak and strong coupling.  These curves are overlaid in black on the experimental data in Fig.~\ref{fig:strong_coupling}C\cite{cohen_supplementary_2022}.

Fig.~\ref{fig:strong_coupling}D shows $G(V = 0, T)$ plotted as a function of the probe temperature for the same values of $V_{\text{NS}}$ as in Fig.~\ref{fig:strong_coupling}C, along with the results of Eq.~\eqref{eq:integral_conductance} for the corresponding values of $T_0$.  
While Eq.~\eqref{eq:integral_conductance} does not provide a simple scaling between temperature and voltage as is available at the weak coupling fixed point, for $V = 0$ the conductance can be written as a function of the scaled temperature, $T/T_0$. 
Fig.~\ref{fig:strong_coupling}E shows the $V=0$ conductance plotted as a function of $T/T_0$.
The four data sets shown in unscaled form in Fig.~\ref{fig:strong_coupling}E collapse onto different parts of the universal crossover function between weak and strong coupling.  Note that for panels D-E, the temperature is measured on the probe and no corrections are made for disequilibrium between $T_\textrm{electron}$ and $T_{\textrm{probe}}$, leading to systematic deviations between  experiment and theory at the lowest temperatures. The collapse of the experimental data onto a single universal curve over nearly two orders of magnitude in $T_0$ strongly supports the conclusion that our quantum point contact realizes the exactly-solvable Lagrangian of Eq.~\eqref{eq:lagrangian} to a high degree of accuracy, and in particular, that tunneling is dominated by a single saddle point in the QPC.

\begin{figure} [ht!]
    \includegraphics[width = 88mm]{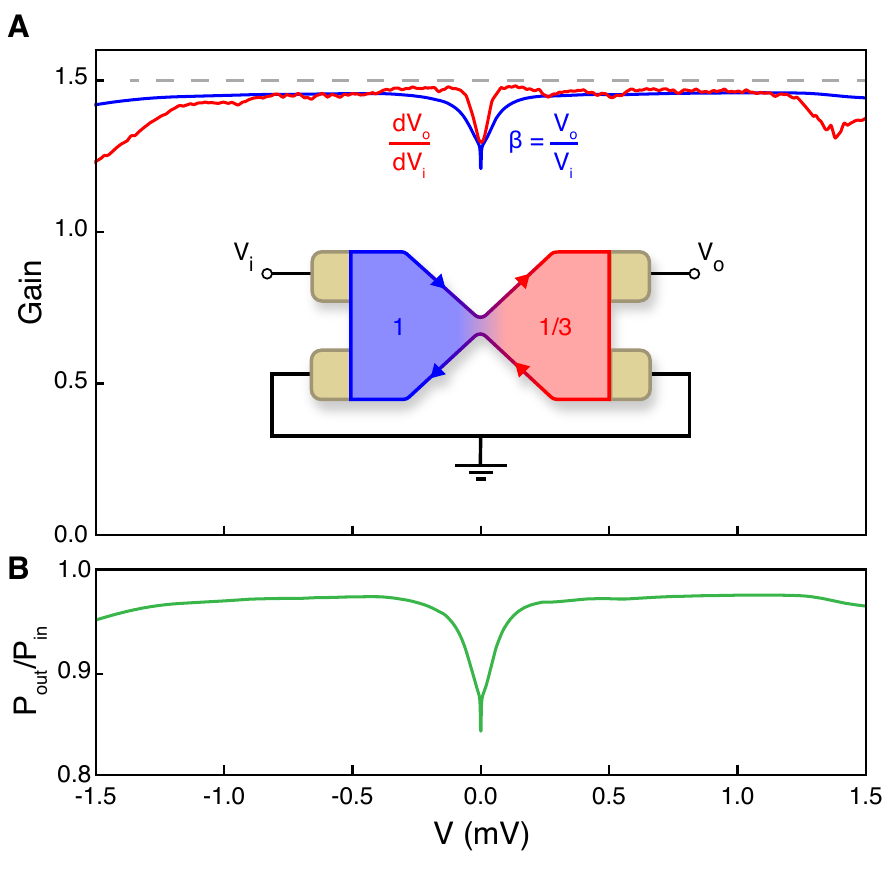}
    \caption{
    \textbf{Zero frequency voltage step-up transformer. }
    \textbf{(a)} The differential gain $dV_\mathrm{o}/dV_\mathrm{i}$ and resulting integrated DC gain $\beta = V_\mathrm{o}/V_\mathrm{i}$, measured in the configuration shown in the inset, with B=9T, $T_{\text{probe}}$=48mK, $V_\mathrm{E}$=-1.460V, $V_\mathrm{W}$=-1.775V, $V_\mathrm{NS}$=-3.225V, $V_\mathrm{BG}$=2.0V. The FQH Andreev scattering process yields an enhancement of the output voltage on the FQH side \cite{chklovskii_consequences_1998}, with the DC gain predicted to reach a value of 1.5 in the dissipationless limit. Experimentally, we find a gain $\beta = 1.46$, despite the nonlinearity at low bias arising from the instability suppression of the  Andreev scattering at low energies.
    \textbf{(b)} The DC power dissipation ratio, calculated from $\beta$ via $P_\mathrm{out}/P_\mathrm{in} = (2 \beta / 3)^2 - (2 \beta / 3) + 1$, is plotted versus V, and reaches a maximum value of 97.6\% \cite{halperin_dc_2003}.
    \label{fig:transformer}
    }
\end{figure}

At the strong coupling fixed point the quantum point contact acts as a nearly dissipationless  splitter, partitioning current injected on the $\nu=1$ edge equally between the downstream $\nu=1/3$ and the upstream $\nu=1$ edge states\cite{sandler_andreev_1998}.  For $\Gamma=\infty$, the partitioning happens with unit probability. In this limit, no entropy is generated and dissipation does not occur, leading to unity power efficiency\cite{chklovskii_consequences_1998}. This contrasts with the more conventional case of partial transmission of edge modes at a quantum point contact, where fluctuations arising from partition noise lead to dissipation. 
The current splitting property of the strong coupling fixed point allows us to operate our device as a voltage step-up transformer\cite{chklovskii_consequences_1998, sandler_andreev_1998}. 
Fig.~\ref{fig:transformer} shows the differential gain, $\mathrm{d} V_o / \mathrm{d} V_i$, where $V_i$ is the input voltage applied to the upstream $\nu=1$ edge state and $V_{o}$ is the output voltage measured on the downstream $\nu=1/3$ edge state with all other contacts grounded.  The differential gain is very close the theoretically expected value of $3/2$ \cite{chklovskii_consequences_1998}.  Remarkably, because it is built on a purely zero frequency effect, the transformer gain remains almost the same even in the `direct current' (DC) limit, as can be seen in the behavior of the integrated gain, $V_o/V_i$, in comparison to its differential counterpart.  The exceptionally high power efficiency, which peaks at $97.6\%$, is a testament to how closely the experiment realizes the strong-coupling fixed point, and contrasts favorably with zero frequency voltage amplification based on superconductors\cite{giaever_dc_1966,girvin_dc_2002} and bilayer quantum Hall systems\cite{girvin_dc_2002, halperin_dc_2003}.

Our observation of universal chiral Luttinger liquid physics at both weak- and strong-coupling directly paves the way for experiments on two dimensional systems where mesoscopic electrostatic control plays a key role in addressing unanswered questions about strong correlations, topological order, and quantum statistics.  Examples include even denominator fractional quantum Hall states observed in mono-\cite{zibrov_even-denominator_2018,kim_even_2019} and bilayer \cite{ki_observation_2014, zibrov_tunable_2017, li_even_2017} graphene, where taming edge reconstruction in a quantum point contact may allow for unambiguous experimental constraints on the ground state topological order \cite{wen_topological_1993,fendley_dynamical_2006,ohashi_andreev-like_2022, bishara_edge_2008, milovanovic_edge_1996}. In addition, the flexibility inherent in anodic oxidation prepatterning will also allow independent tuning of quasiparticle number, edge sharpness, and quantum point contact transparency in interferometer geometries where direct access to quantum statistics are possible\cite{nayak_non-abelian_2008,nakamura_aharonovbohm_2019}.  Finally, gate-defined point contacts may allow for precision measurements of order parameters in the recently discovered crystalline graphene superconductors\cite{zhou_superconductivity_2021,zhou_isospin_2022,zhang_spin-orbit_2022}.  


\section*{Data Availability}
The data that support the findings of this study are available from the corresponding author upon reasonable request.

\section*{Acknowledgements}
The authors acknowledge discussions with B. L. Halperin, C. Nayak, and Zhenghan Wang, as well as A. Assouline and A.M. Potts for helpful comments on the manuscript.   
Work at UCSB was primarily supported by the Air Force Office of Scientific Research under award FA9560-20-1-0208 and by the Gordon and Betty Moore Foundation EPIQS program under award GBMF9471. 
LC and NS received additional support from the Army Research Office under award W911NF20-1-0082. 
TW and MZ were supported by the Director, Office of Science, Office of Basic Energy Sciences, Materials Sciences and Engineering Division of the U.S. Department of Energy under contract no. DE-AC02-05-CH11231 (van der Waals heterostructures program, KCWF16).
MZ received additional support from the Army Research Office through the MURI program (grant number W911NF-17-1-0323)
K.W. and T.T. acknowledge support from JSPS KAKENHI (Grant Numbers 19H05790, 20H00354 and 21H05233).

\let\oldaddcontentsline\addcontentsline
\renewcommand{\addcontentsline}[3]{}
\bibliographystyle{unsrt}
\bibliography{references}
\let\addcontentsline\oldaddcontentsline

\clearpage
\pagebreak

%


\title{Supplementary Information}
\maketitle
\onecolumngrid

\renewcommand\thefigure{S\arabic{figure}}
\setcounter{figure}{0}

\section{Junction Electrostatics: Reconstruction Effects and Resonant Scattering}

\begin{figure}[b!]
    \includegraphics[width = 179mm]{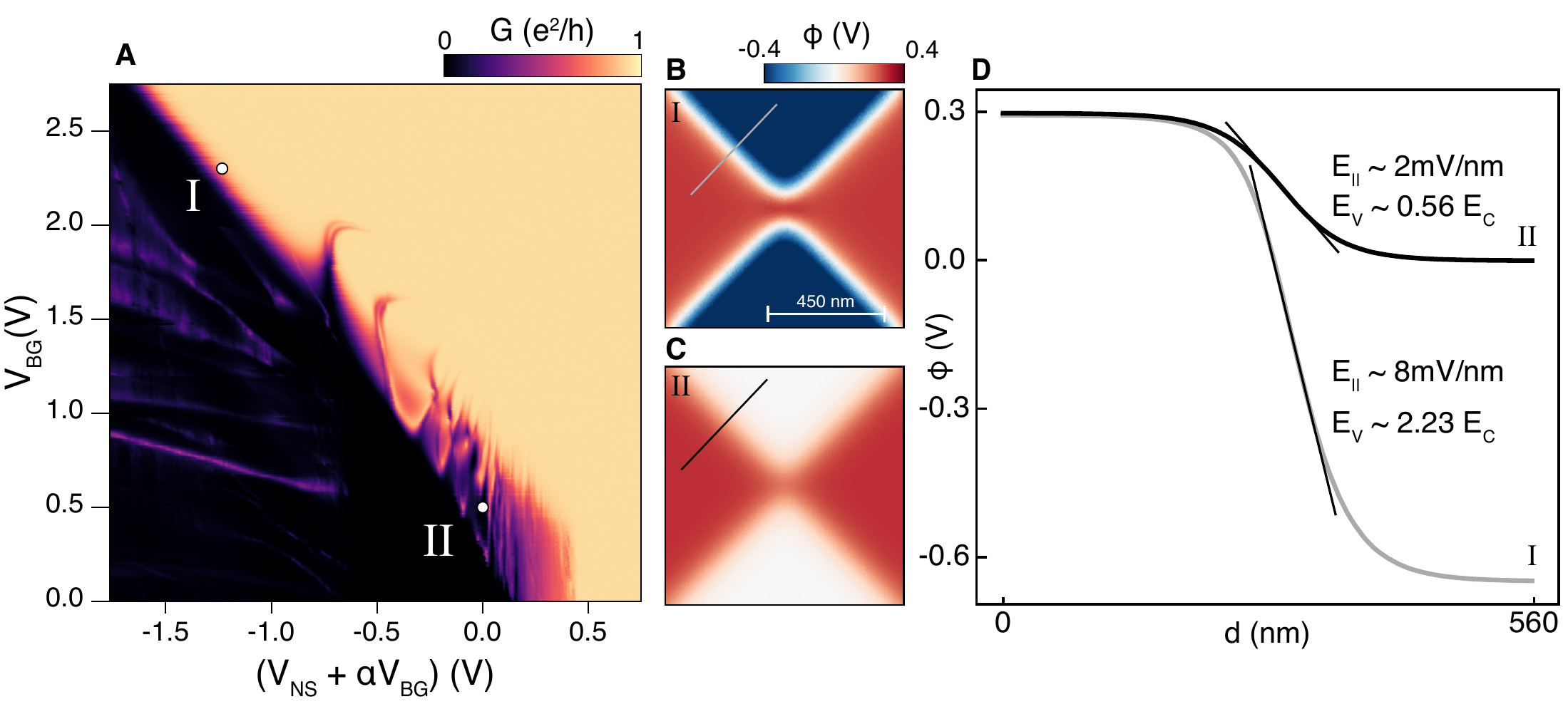}
    \caption{
    \textbf{Signatures of reconstruction in $G$ between $\nu = 1$ edge modes.} 
    \textbf{(a)} The conductance measured across the QPC with both the East and West regions in $\nu = 1$ at B = 8T. The East and West gate voltages are swept in the opposition direction of $V_{\text{BG}}$ along the y axis, through the range $V_\mathrm{EW} \in (0.6V, -2.191V)$, to maintain a fixed filling factor while varying the voltage difference and thereby the potential sharpness.
    \textbf{(b)} The simulated electric potential at the monolayer, corresponding to the operating point I.
    \textbf{(c)} Same as (b) but at the operating point II, where the potential is much softer.
    \textbf{(d)} Simulated potential along the contours marked in grey and black in panels B and C, respectively. The softness is quantified by the maximum magnitude of the in-plane confining electric field, $E_{\parallel}$ (\textit{i.e.} simply the gradient of the potential normal to the boundary between the N(/S) and E(/W) regions).
    \label{fig:electrostatics_integer}
    }
\end{figure}

\begin{figure}[ht!]
    \includegraphics[width = 179mm]{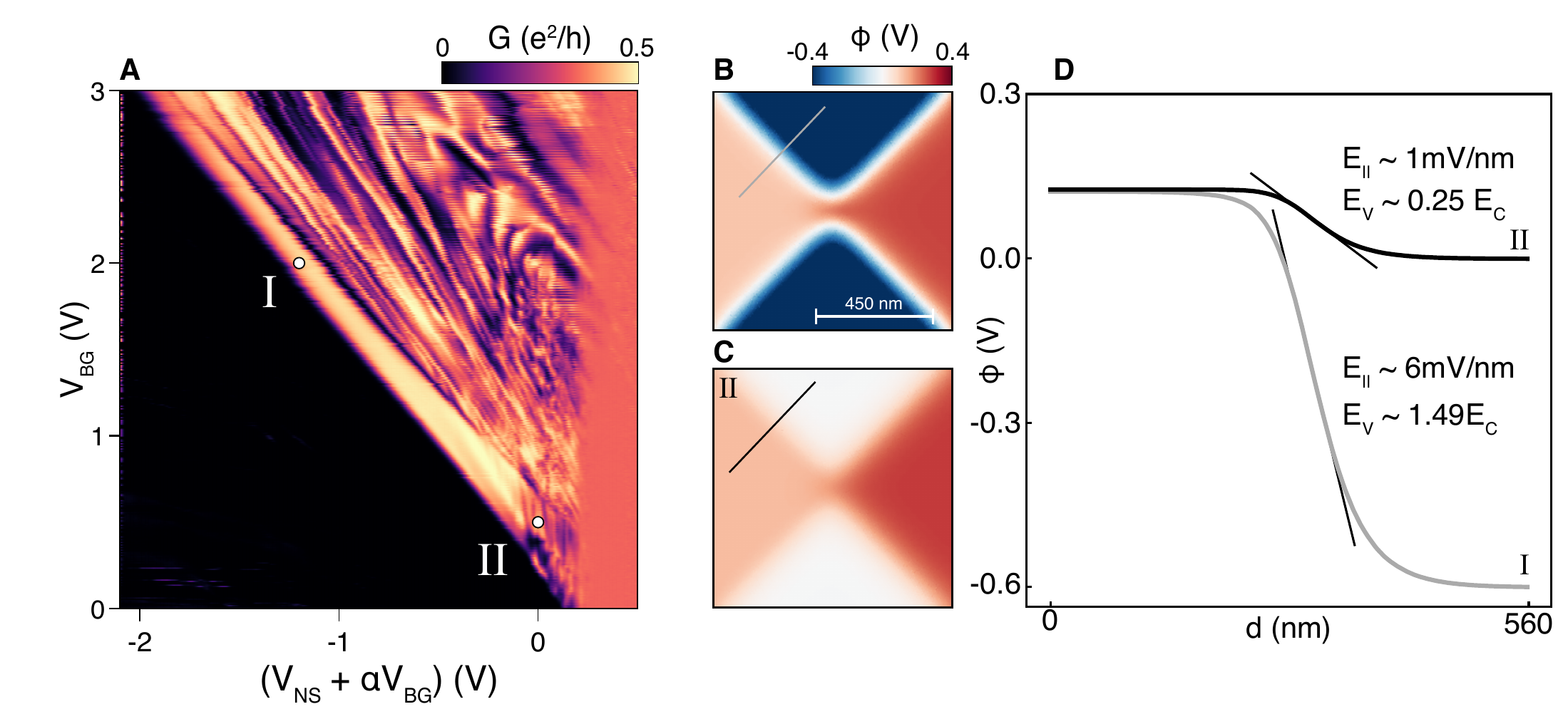}
    \caption{
    \textbf{Signatures of reconstruction in $G$ between $\nu = 1$ and $\nu = 1/3$ edge modes} 
    \textbf{(a)} The conductance measured across the QPC with $\nu_\mathrm{W} = 1/3$ and $\nu_\mathrm{E} = 1$ at B = 9T. Along the y-axis, both the East and West gates must be swept in the opposite direction of the back gate, but at distinct voltages $V_\mathrm{E} \in (0.57V, -2.475V), V_\mathrm{W} \in (0.256V, -2.79V)$, to maintain fixed filling factors.
    \textbf{(b)} The simulated electric potential at the monolayer, corresponding to the operating point I.
    \textbf{(c)} Same as (b) but at the operating point II, where the potential is much softer.
    \textbf{(d)} Simulated potential along the contours marked in grey and black in panels B and C, respectively, at the boundary of the $\nu_\mathrm{W}=1/3$ region. The boundary of the $\nu_\mathrm{E}=1$ region is necessarily sharper since $V_\mathrm{E}-V_\mathrm{BG} > V_\mathrm{W}-V_\mathrm{BG}$.
    \label{fig:electrostatics_fractional}
    }
\end{figure}
If the slope of the confining electric potential is soft a bevy of edge reconstruction effects may arise.  
These include resonant transmission through reconstruction-stabilized local states, complicating the interpretation of the conductance across the QPC. 
In this manuscript we focus on the regime of ``sharp'' electrostatics where the edge reconstructions are negligible, which can be achieved by tuning the gates in our device geometry\cite{cohen_tunable_2022}. 
The sharpness of a potential is defined via the competition between two energy scales: the ``confinement energy'' associated with the applied potential, $E_V \equiv e | \nabla V | \ell_B$, and the Coulomb energy, $E_C = \frac{e^2}{4\pi \epsilon \ell_B}$.  
$E_V < E_C$ corresponds to the ``soft'' electrostatic regime; here, the charge may distribute itself primarily by minimizing the Coulomb repulsion energy.  Along a translation-invariant quantum Hall edge, this leads to the formation of additional incompressible strips along the boundary, resulting in additional edge modes that do not  correspond to the primary bulk filling factor\cite{chamon_sharp_1994, khanna_fractional_2021}.  
In quantum point contacts made identically to the one presented in this work, the soft-edge regime is characterized by local islands of charge stabilized at the potential saddle point, which mediate resonant tunneling between edge modes across the quantum point contact\cite{cohen_tunable_2022}.  
In contrast, in the sharp electrostatic regime where the confinement energy is larger than the Coulomb energy, $E_V/E_C > 1$, simple tunneling between two edge modes at a point is recovered.

The gate voltages necessary to achieve appropriately sharp confinement can be identified by observing the qualitative behavior of the conductance near pinch-off. First we examine a regime where the phenomenology is simplest---Fig.~\ref{fig:electrostatics_integer}A presents a plot of the conductance measured with a fixed filling factor $\nu = 1$ in both the East and West quadrants. 
On the x-axis, $V_\mathrm{NS}$ is changed to vary the filling factor in the North and South quadrants across a fixed range of densities by changing the quantity $V_\mathrm{NS} + \alpha V_\mathrm{BG}$, where $\alpha$ is the capacitance ratio between the top and bottom graphite gates. Along the y axis, $V_\mathrm{BG}$ is swept while concurrently sweeping $V_\mathrm{EW}$ in the opposite direction to fix the quantity $V_\mathrm{EW} + \alpha V_\mathrm{BG}$, maintaining a constant filling factor in the East and West quadrants while varying the difference $V_\mathrm{EW} - V_\mathrm{BG}$, which effectively tunes the confinement energy $E_V$.

To illustrate this in more detail, we analyze the electrostatics near the points marked I and II in Fig. \ref{fig:electrostatics_integer}A, which are representative of the ``sharp'' and ``soft'' regimes, respectively. 
In the ``sharp'' regime represented by point I, the conductance exhibits a monotonic step from $G=0$ to $G=\frac{e^2}{h}$ as the QPC is opened; this is what is expected for non-resonant tunneling between two integer edge modes at a QPC. In contrast, for the ``soft'' regime represented by point II, the conductance behaves non-monotonically, displaying several peaks and dips as the junction is opened. 

Fig.~\ref{fig:electrostatics_integer}B-C present finite element analysis simulations of the electrostatic potential at the graphene layer of the device (calculated using COMSOL Multiphysics) for an idealized device geometry with voltages corresponding to these two operating points. Fig.~\ref{fig:electrostatics_integer}D shows the potential along the illustrated lines normal to the boundary between the $\nu = 1$ and depleted regions.
We can quantitatively characterize the threshold sharpness by estimating the ratio $E_V/E_C$ from the simulated confining electric field. At $B=8$~T the Coulomb energy is $E_C \approx 33$~meV, where the relevant $\epsilon=\sqrt{\epsilon_\perp \epsilon_\parallel}\approx4.8$ taking $\epsilon_\perp=3.5$ and $\epsilon_\parallel=6.6$. 
Experimentally, the effects of resonant transmission disappear above $E_V \approx 61$~meV, so $E_V/E_C > 1.85$.   

Fig.~\ref{fig:electrostatics_integer}A makes it clear that in the IQH regime, local effects from edge reconstruction can be suppressed by making the confining potential sufficiently sharp. 
To find a similarly universal regime for the $\nu=1$ to $\nu = 1/3$ configuration, we repeat the characterization above with the East and West regions fixed to those filling factors, shown in Fig.~\ref{fig:electrostatics_fractional}A. While many additional peaks and dips occur beyond the initial conductance step, many features are consistent with the $\nu=1$ to $\nu=1$ case. First, focusing only on the region near the first conductance step, when the QPC is just barely open, the qualitative features are similar Fig.~\ref{fig:electrostatics_integer}A, with a smooth step up to a plateau (here with a value of $G \approx \frac{e^2}{2h}$) in the sharp regime, near point I, and number of resonant peaks arising in the soft regime at point II.
Notably, resonant transmission effects near the initial conductance step  disappear at a threshold of about $E_V > 30$~meV, indicating a much lower threshold than in the IQH case. The data sets presented in the main text are measured at relatively sharp confinement, similar to that of point I for the $B=9$~T and $B=10$~T data.

\begin{figure*}
    \includegraphics[width = 179mm]{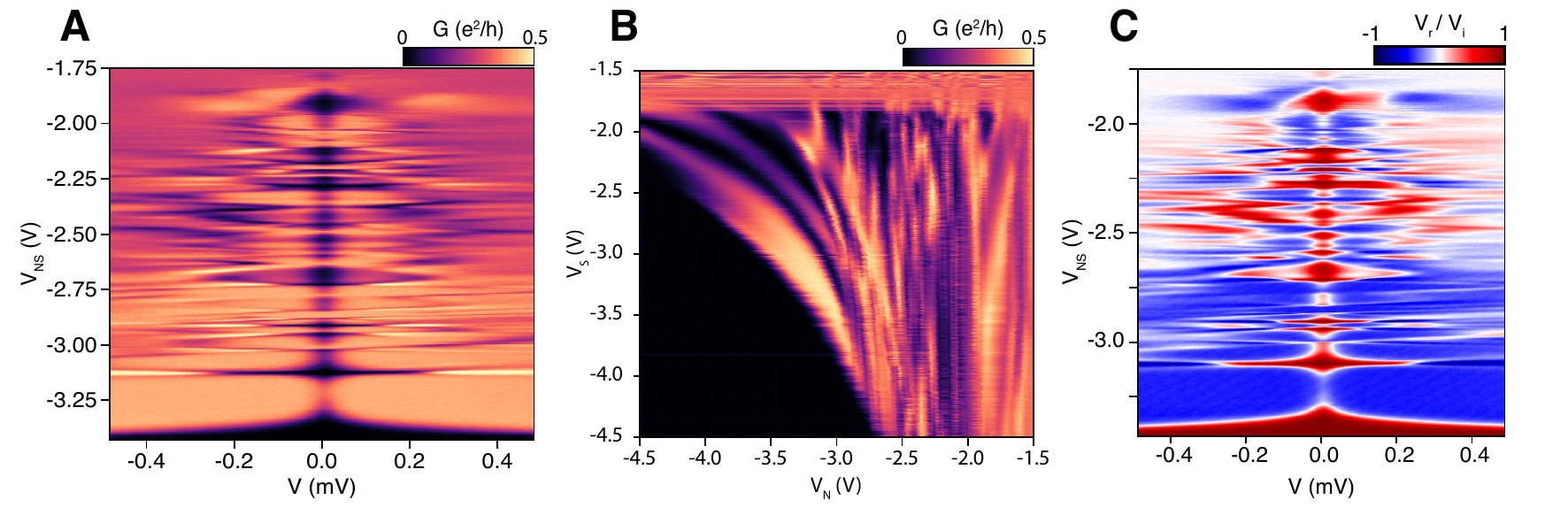}
    \caption{
    \textbf{Resonances at high QPC transmission.} 
    \textbf{(a)} The differential  conductance across the $1$-$1/3$ junction vs. North/South gate voltage $V_\mathrm{NS}$ and DC bias voltage $V$: B=9T, T=56mK, $V_\mathrm{E}$=-1.460V, $V_\mathrm{W}$=-1.775V, $V_\mathrm{BG}$=2.0V. A large number of resonant dips and peaks arise as the voltage $V_\mathrm{NS}$ is increased, effectively widening the junction.
    \textbf{(b)} The zero-bias conductance $G$ in the same regime, but varying $V_{\text{N}}$ and $V_{\text{S}}$ independently.  Following the rationale of reference \cite{cohen_tunable_2022} we may determine the ratio of the capacitance of the local island to the North and South gates from the slope of the resonant peak trajectory: $\frac{dV_\mathrm{N}}{dV_\mathrm{S}} = - \frac{C_\mathrm{S}}{C_\mathrm{N}}$. \textbf{(c)} The ratio of the reflected voltage to the incident voltage, $V_r / V_i$, plotted against the DC voltage bias $V$, and the North/South gate voltages $V_{\text{NS}}$ under the same conditions as panel A.  
    \label{fig:resonances}
    }
\end{figure*}

The most obvious difference between Figures \ref{fig:electrostatics_integer}A and \ref{fig:electrostatics_fractional}A is the sheer number of features visible within the range beyond the initial conductance step, where the QPC is very ``open.'' 
The presence of nontrivial structure in this regime is due to a fundamental difference between the $\nu = 1$ to $\nu = 1$ junction and the $\nu = 1$ to $\nu = 1/3$ heterojunction, with the latter involving a change in the topological order of the bulk. To explore this further, Fig.~\ref{fig:resonances}A shows the differential conductance as a function of both the DC bias $V$ and $V_\mathrm{NS}$ starting near the operating point marked I in Fig.~\ref{fig:electrostatics_fractional}A. The data in the main text focuses on the regime near the first step (analogous to $V_{\text{NS}} < -3.25V$ in Fig.\ref{fig:resonances}A). However, extending the range of $V_{\text{NS}}$ reveals additional structure.  Throughout the plot, the conductance dips nearly to $0$ several times near $V = 0$. 
Notably, the first several low-bias dips are reminiscent of the initial conductance step analyzed in the main text, where the nonlinearity results from the chiral Luttinger liquid nature of the $\nu = 1/3$ edge. An appealing narrative is that at least the first several resonances arise intrinsically, from reconstruction within the point contact, resulting in a single re-entrant resonant scatterer with similarly universal physics\cite{kane_resonant_1998}.

We may test this idea by independently varying $V_\mathrm{N}$ and $V_{\text{S}}$, shown in Fig.\ref{fig:resonances}B.  
From this data, it is evident that the first several features follow curved trajectories in $V_\mathrm{N}$ vs. $V_\mathrm{S}$, as expected for a scattering center that varies its position strongly with gate voltages.  This is consistent with an intrinsic effect tied to reconstruction, where the scatterer is pinned to the saddle point in the applied potential (see \onlinecite{cohen_tunable_2022} for a similar analysis). 
To further explore the role of resonant scattering, Fig.~\ref{fig:resonances}C shows the voltage of the reflected $\nu = 1/3$ edge, measured in the same configuration shown in Fig.~\ref{fig:resonances}A.  $V_r/V_i$ attains a negative value at finite bias even when the low-bias conductance is suppressed, both on the initial conductance step discussed in the main text but also between the first few resonances; indeed after the first resonance a nearly quantized $V_r/V_i\approx-1/2$ is again observed. 

As the QPC is opened further, additional conductance dips appear accompanied by their own strong nonlinearities.  
For $V_{NS}\gtrsim -2.75V$, these features show little curvature in $V_\mathrm{N}-V_\mathrm{S}$, and seem to depend more strongly on $V_{\text{N}}$ than $V_{\text{S}}$. This is consistent with scattering mediated by extrinsic disorder centers that are fixed in space. Concomitant with the onset of these features, the conductance at high bias tends toward $G = e^2/3h$ (rather than $e^2/(2h)$), and $V_r$ becomes positive even at high bias. 

The  behavior we observe for the open junction is qualitatively consistent with theoretical models, such as those proposed in Ref. \onlinecite{chamon_distinct_1997}. In the integer quantum Hall case (Fig. \ref{eq:integral_conductance}), a fully open QPC is equivalent to a uniformly doped macroscopic sample: there is no communication between chiral edge states on opposite edges, and the conductance is quantized at $G = \frac{e^2}{h}$ for arbitrary junction width.  In contrast, in the fractional case of Figs. \ref{fig:electrostatics_fractional} and \ref{fig:resonances}, opening the QPC more widely simply leads to an ``edge state'' following a vertical trajectory through the center of the QPC. Equilibration between the integer and fractional quantum Hall edge states is necessarily going to be sensitive to the presence of localized scatterers along this boundary.  In the wide junction limit, the presence of multiple, possibly coherent scattering processes complicates analysis of the tunneling process, restoring the `line junction' limit first explored by Refs. \onlinecite{chang_observation_1996,chang_chiral_2003}.



\section{Data Analysis}

\subsection{Effects of cryogenic electronic filters}
The measurements presented in this paper were performed in a dilution refrigerator setup with a base temperature of $\SI{56}{mK}$.  To improve thermalization of the electron system to the phonon bath, we heavily filter the measurement wires.  However, these filters give a finite contribution to the measured value of $G$ that must be corrected for to give quantitatively reliable results.  In this section we describe how the true conducatance is extracted from the measured voltages and currents at room temperature. 

\begin{figure*}[b!]
    \includegraphics[width = 179mm]{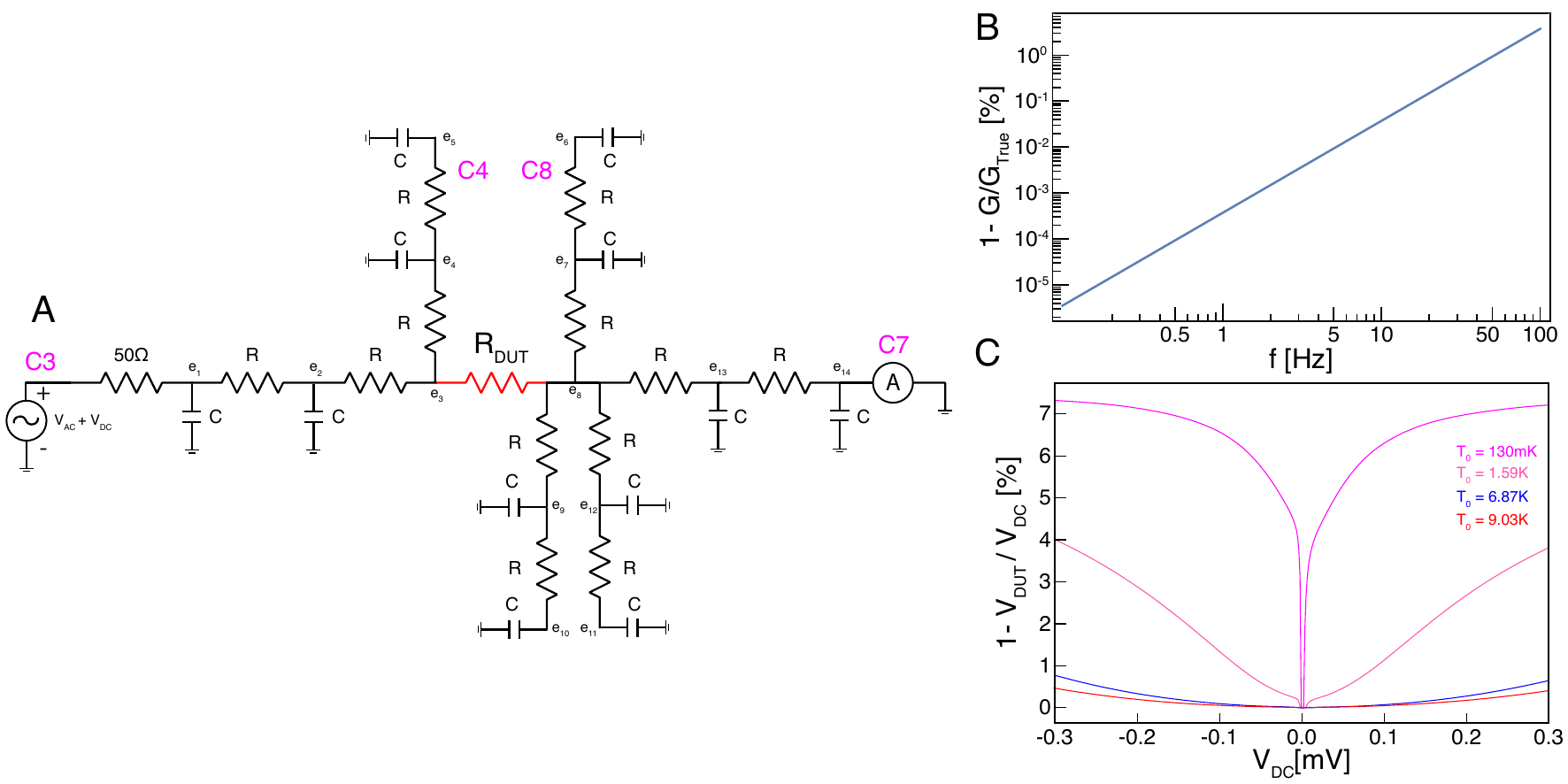}
    \caption{
    \textbf{Measurement circuit including electronic filters.} 
    \textbf{(a)} Model of the sample plus the electronic low-pass filters attached to each contact.  Here $R = 1 k\Omega$ and $C = 20$nF.  \textbf{(b)} Percentage deviation of the measured value of the conductance $G \equiv \text{Re}[e_{13}] / (R \cdot \text{Re}[e_5 - e_6])$ to the true value of the conductance $G_{\text{true}}$. \textbf{(c)} Percentage deviation of the corrected bias voltage ($V_{\text{DUT}})$ from the applied bias voltage ($V_{\text{DC}})$ versus $V_{\text{DC}}$.  The value of this deviation is plotted for each of the curves given in the main text Fig. \ref{fig:strong_coupling}C.
    }
    \label{fig:filter_schematic}
\end{figure*}

All measurements were performed using standard lock-in techniques with an excitation frequency of $f = \SI{2.74}{Hz}$.  
In order to extract the differential tunneling conductance we measure $G \equiv I / V_D$, shown schematically in Fig. \ref{fig:fig1}B.  
We source current via C3 on the $\nu = 1/3$ side of the junction and measure the output current at C7 through the Ithaco 1211 transimpedance amplifier.  The diagonal voltage $V_D$ is extracted by measuring the voltage drop between C4 and C8 with an SR560.  The Ithaco is modeled as a perfect ammeter, while the SR560 is treated as a perfect voltmeter.  Furthermore, we assume -- due to the chiral nature of the edge states -- that upstream contacts from the source do not contribute to the cross-device conductance.  The equivalent device circuit schematic is given in Fig. \ref{fig:filter_schematic}A.  Each device contact is effectively connected to room temperature instrumentation, or ground, via a two-stage low-pass RC filter; while there are RF low-pass Pi filters, they have small component values and can be ignored in the low-frequency limit.  While only C3 and C4 contribute on the $\nu = 1/3$ side of the junction (C2/C1 are upstream), C5/C6 are downstream of the source and must be included as floating contacts on the $\nu = 1$ side of the junction.  Tunneling at the heterojunction is modeled as a resistor labeled $R_{\text{DUT}}$ which should be considered as voltage bias dependent. 

Given the extra components surrounding $R_{DUT}$, it is not guaranteed for every frequency that an AC four-terminal measurement will provide an accurate measure of the true differential conductance, $G_{\text{true}}$.  Using the notation defined in Fig. \ref{fig:filter_schematic}A, $V_D \equiv \text{Re}[e_6 - e_5]$,  and $I \equiv \text{Re}[(e_{13} - e_{14})] / R \approx \text{Re}[e_{13}]/R$.  At our measurement frequency of $f = \SI{2.74}{Hz}$, it can be shown by solving the set of Kirchoff's laws -- set by the schematic in Fig. \ref{fig:filter_schematic}A -- that $G \equiv I/V_D = \text{Re}[e_{13}] / (R \cdot \text{Re}[e_6 - e_5])$ is a good approximation to $G_{\text{true}}$. The percentage error, $\eta \equiv 100 \frac{G_{\text{true}} - G }{G_\text{true}}$, is plotted for a fixed value of $G_{\text{true}} = 10^{-3} e^2/h$ in Fig. \ref{fig:filter_schematic}B versus frequency.  While $G \approx G_{\text{true}}$ for sufficiently low frequencies, $\eta$ scales as $f^2$, and quickly grows to be large; additionally $\eta$ decays monotonically with increasing $G_{\text{true}}$ so the value of $\eta$ with $G_{\text{true}} = 10^{-3} e^2/h$ represents a good upper-bound on the error given the resistance ranges in our measurement.  For $f > \SI{75}{Hz}$, $\eta$ becomes greater than $1\%$ even for values of $G_{\text{true}} \gg e^2/h$.   Consequently, we chose to measure at a frequency significantly lower than this threshold such that any reduction in $f$ resulted in an experimentally undetectable change in $G$, and so that we could safely assume $G \approx G_{\text{true}}$ over a wide range of conductance values.

While at $f = \SI{2.74}{Hz}$, four-terminal measurements of $G$ remain a good approximation of $G_{\text{true}}$, the DC voltage bias across $R_{\text{DUT}}$ requires some additional analysis.  In order to investigate the DC bias directly across the sample it is necessary to take into account the filter resistances since they inherently form a voltage divider between themselves and $R_{\text{DUT}}$. Determining the relationship between the applied DC voltage at the input of the filters (a known value defined as $V_{\text{DC}}$) and the voltage bias across the sample, defined as $V_{\text{DUT}}$, is straightforward.  At low frequencies, the current drawn to the floating contacts is negligible so the DC voltage across the sample is well approximated by 
\begin{equation}
    V_{\text{DUT}} = V_{\text{DC}} - (4R + 50\Omega) \cdot I
\end{equation}
where $I$ is the DC current through $R_{\text{DUT}}$.  While the DC current is not known a priori, differentiating with respect to $V_{\text{DUT}}$ the above equation gives:
\begin{equation}
    1 = \frac{\mathrm{d} V_{\text{DC}}}{\mathrm{d} V_{\text{DUT}}} - (4R + 50\Omega) \frac{\mathrm{d} I}{\mathrm{d} V_{\text{DUT}}}
\end{equation}
This equation can be rearranged to isolate $\mathrm{d} V_{\text{DUT}}$ and $\mathrm{d} V_{\text{DC}}$, given below.
\begin{equation}
    \mathrm{d} V_{\text{DUT}} = \frac{\mathrm{d} V_{\text{DC}}}{1 + (4R + 50\Omega) \frac{\mathrm{d} I}{\mathrm{d} V_{\text{DUT}}}} 
\end{equation}
The differential conductance, $G_{\text{true}} = \mathrm{d} I/ \mathrm{d} V_{\text{DUT}} \approx G$, appears in the denominator.  Integrating this equation leaves the final correction formula, which can be expanded for $4050 \Omega \cdot G \ll 1$ to give:
\begin{equation}
    V_{\text{DUT}} =  \int_0^{V_{\text{DC}}} \frac{\mathrm{d} V_{\text{DC}}'}{1 + (4R + 50\Omega) G_{\text{true}} (V_{\text{DC}}')} \approx V_{\text{DC}} - (4050\Omega) \int_0^{V_{\text{DC}}} \mathrm{d} V_{\text{DC}}' G(V_{\text{DC}}')
\end{equation}

While in general this correction remains small for low conductance values, indicating the bias voltage in the range where universal power-law behavior is observed is unaffected, for measurements where the conductance is large for the entire bias-dependence the corrections can be significant.  The effect of this correction is plotted in Fig. \ref{fig:filter_schematic}C corresponding to the tunneling data presented in Fig. \ref{fig:strong_coupling}C.  For the largest values of $T_0$, i.e, $T_0 = 
\SI{6.87}{K}$ and $T_0 = \SI{9.02}{K}$ respectively, the corrected bias remains less than $1\%$ of the applied bias voltage.  However, for lower values of $T_0$ this correction becomes significant.

\subsection{Determining the theoretically expected range of universal power-law scaling}
For sufficiently low temperatures and bias voltages, the Lagrangian in Eq.~\eqref{eq:lagrangian} is expected to flow to the weakly coupled fixed point where $G \propto V^2, T^2$.  However, while in this regime where the temperature and bias dependence is expected to be universal and irreverent of microscopic details, the range over where this behavior persists does care about the sample specifics. Even in the most ideal case, where Eq.~\eqref{eq:lagrangian} describes the system for \textit{all} bias voltages and temperatures, the range where the $O(V^2, T^2)$ terms in Eq.~\eqref{eq:lutt_cond} dominate is determined by the bare value of $\Gamma$.  As such, when extracting the power-law exponents and assessing the quality of the scaling collapse it is important to not over-extend the fitting range beyond where universality is expected, i.e, where $G \propto V^2, T^2$.  In general, while the cross over from weak to strong coupling predicts deviations from $G \propto V^2, T^2$ behavior at sufficiently high energies, additional perturbations in Eq.~\eqref{eq:lagrangian} will generally reduce the range where universal behavior may be expected.  Consequently, the bias voltages and temperatures given by the quantum impurity model where significant deviations from $G \propto V^2, T^2$ are predicted provides a self-consistent range where power law behavior should be observed. 

For our fridge base temperature of $T_{\text{probe}} = \SI{56}{mK}$, we may evaluate the threshold bias voltage $V_{\text{Th}}$ above which the conductivity predicted by Eq.~\eqref{eq:lutt_cond} exceeds the conductivity given by the full quantum impurity model (Eq.~\eqref{eq:integral_conductance}) by more than $5\%$.  For a simulated temperature of $T = \SI{56}{mK}$, when $T_0 = \SI{9}{K}$, $V_{\text{Th}} \approx 170\mu V$.  Consequently, in order to extract $T_0$ for the data presented in Fig. \ref{fig:fig1}C, $T_0$ is initially guessed to be $\SI{9}{K}$.  $T_0$ is then fit over a range of $V = \SI{15}{\mu V}$ to $V = \SI{170}{\mu V}$.  The fit results in a $T_0 = \SI{9.02 \pm 0.007}{K}$, which does not effectively change the fitting range given a bias step of $\SI{5}{\mu V}$ per data point.  This procedure leads to a self-consistent result where both $T_0$ and the range over which $T_0$ is fit simultaneously converge.  

\begin{figure*}[ht]
    \centering
    \includegraphics[width = 179mm]{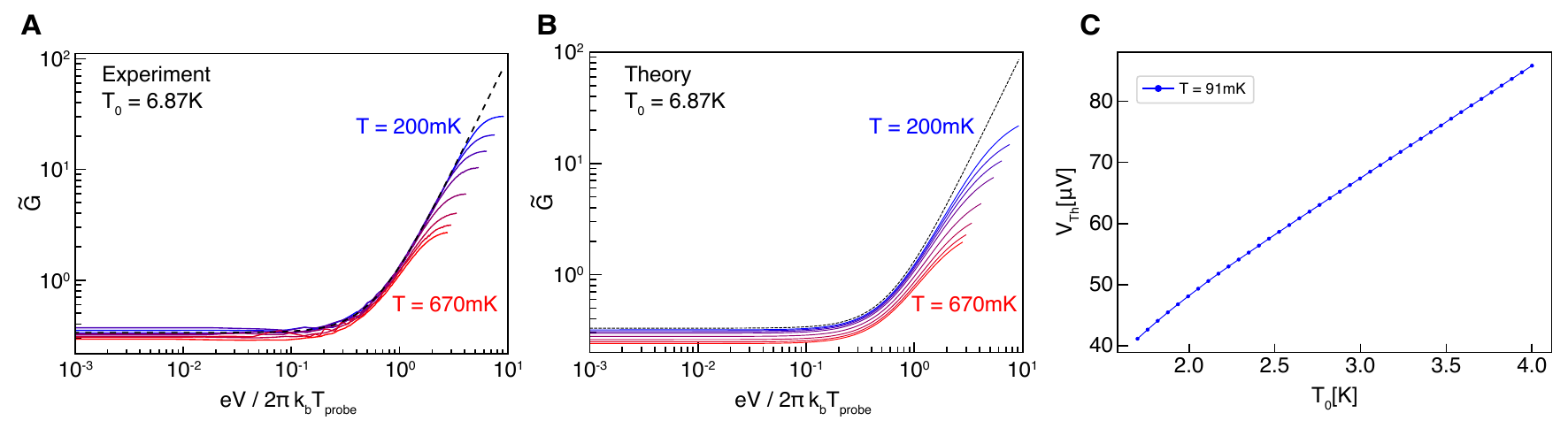}
    \caption{\textbf{Analysis of Deviations from Universal Luttinger Liquid Behavior.} \textbf{(a)} The scaled conductance $\widetilde{G} = \frac{2h}{e^2} (\frac{T_0}{2\pi T})^2 G$ versus the scaled voltage bias $eV / 2\pi k_b T$ as in the main text Fig.\ref{fig:fig1}F (temperature values are listed in the caption of Fig.\ref{fig:fig1}). \textbf{(b)} Theoretically predicted scaled conductance $\widetilde{G}$ versus $eV / 2\pi k_b T$.  Here $\widetilde{G}$ is computed the same way as in panel (a), however, the unscaled conductance is computed from the conductance in Eq. \eqref{eq:integral_conductance} from the full quantum impurity model. Black dashed line is the theoretical curve for the low energy scaled conductance: $\widetilde{G} = 1/3 + x^2$.  $T_0 = 6.87K$ and the values of $T$ are kept the same as in panel (a). \textbf{(c)} The threshold voltage $V_{\text{Th}}$ below which the Sommerfeld expansion of $O(V^2, T^4)$ of  Eq.~\eqref{eq:integral_conductance} (given in Eq.\eqref{eq:sommerfeld}) is expected to be valid for a fixed $T = \SI{91}{mK}$.}  
    \label{fig:range_of_universality}
\end{figure*}

We may also consider the scaling collapse in Fig.\ref{fig:fig1}F, which in particular, is only valid when $T << T_0 / 2\pi$.  The experimental scaling collapse data is repeated in Fig.\ref{fig:range_of_universality}A and may be directly compared to Fig.\ref{fig:range_of_universality}B showing the theoretically predicted scaled conductance for a $T_0 = \SI{6.87}{K}$ for the same temperature values used in the experimental scaling collapse.  The scaling collapse begins to fail under two conditions; $eV >> k_b T$, and $T >> T_0 / 2\pi$. From Fig.\ref{fig:range_of_universality}B it is clear that at low voltage bias, the curves are expected theoretically to collapse on to each other well up to $T = \SI{550}{mK}$, then are predicted to deviate from the low-energy prediction of Eq.\eqref{eq:lutt_cond} at higher temperatures.  Additionally, all the curves regardless of temperature deviate from Eq.\eqref{eq:lutt_cond} at sufficiently high bias.  In direct comparison to Fig.\ref{fig:range_of_universality}B, the the onset of deviations from from Eq.\eqref{eq:lutt_cond} in both temperature and voltage bias in the experimental data in Fig.\ref{fig:range_of_universality}A are well capture by the theoretical model to within experimental uncertainty.

In general, as $T_0 \sim T/2\pi$ for the lowest experimentally accessible temperatures, Eq.~\eqref{eq:lutt_cond} will no longer be an accurate model of the system for any range of voltage bias.  However, it may be observed that the data presented in Fig. \ref{fig:strong_coupling}C with $T_0 = \SI{1.59}{K}$ does have a $V^2$ dependence for some substantial range of $V$ despite $T_0 / 2\pi \approx \SI{250}{mK}$ being sufficiently close to the lowest available electron temperatures.  This can be understood by computing the Sommerfeld expansion of Eq.~\eqref{eq:integral_conductance} to $O(V^2, T^4)$, given in Eq.~\eqref{eq:sommerfeld}.
\begin{equation}
    \label{eq:sommerfeld}
    G(V, T) = \bigg [\frac{1}{6} (\frac{2\pi T}{T_0})^2 - \frac{7}{30} (\frac{2\pi T}{T_0})^4 \bigg] + \bigg [\frac{1}{2} - (\frac{2\pi T}{T_0})^2 + \frac{7}{2}(\frac{2\pi T}{T_0})^4 \bigg] (\frac{eV}{k_b T_0})^2
\end{equation}
$V_{\text{Th}}$, now calculated as the bias voltage above which the conductivity predicted by Eq.~\eqref{eq:sommerfeld} differs from that predicted by Eq.~\eqref{eq:integral_conductance} by more than $5\%$, is given in Fig.~\ref{fig:range_of_universality}B for a fixed simulated temperature of $T = \SI{91}{mK}$ versus $T_0$.  It is clear that even for a moderate temperature of $\SI{91}{mK}$, the range of $V$ where a quadratic bias dependence is expected is significant.  While the conductivity in Eq.~\eqref{eq:sommerfeld} is still entirely parameterized by a single value $T_0$, the curvature of the bias depdence is weakly temperature dependent.  This becomes relevant when attempting to calculate $T_0$ from the low-energy bias dependence for the data presented in Fig.~\ref{fig:strong_coupling}C with $T_0 \sim 1.6K$.

\newpage
\subsection{Extracting power-law exponent}
\begin{figure*}[ht]
    \centering
    \includegraphics[width = 179mm]{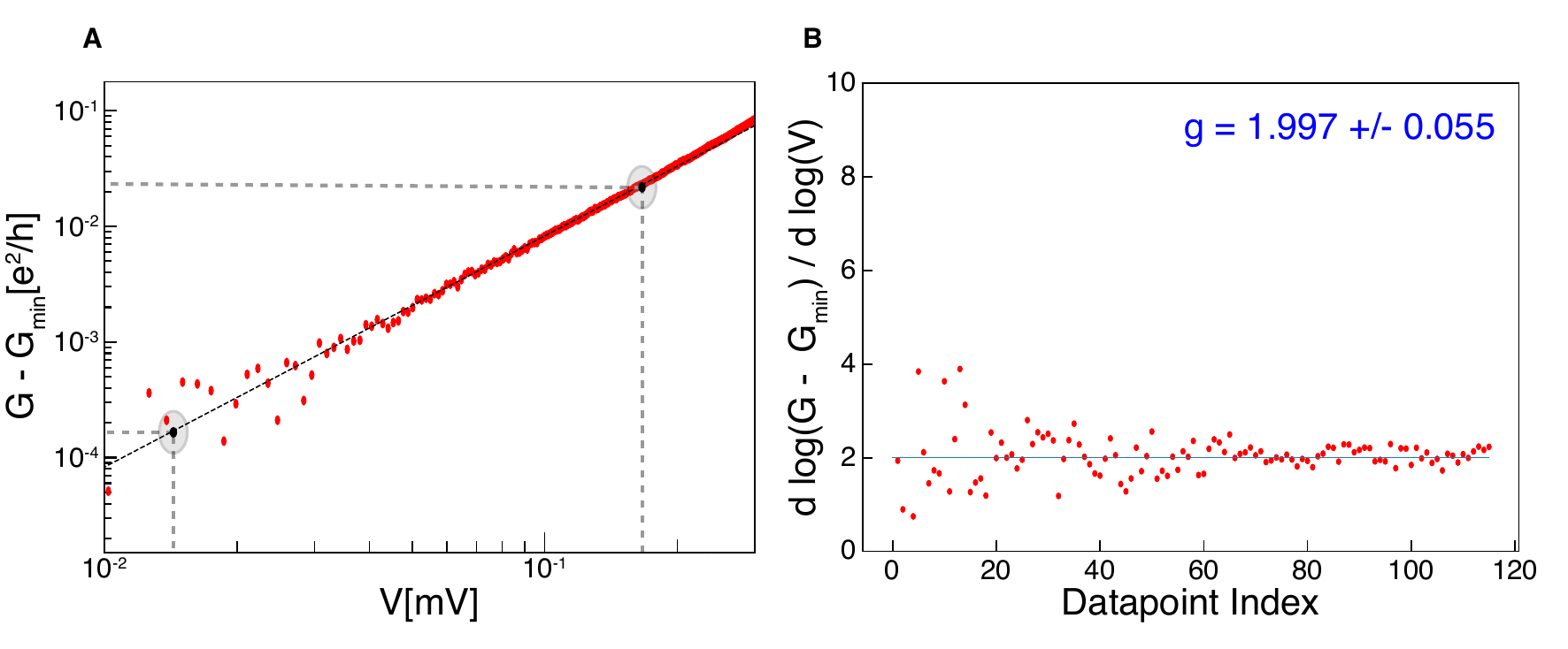}
    \caption{\textbf{Computing power law exponent from the voltage bias-dependence} \textbf{(a)} $G - G_{\text{min}}$ as in Fig \ref{fig:fig1}C.  Points marked by gray circles indicate upper and lower bounds of included data ($V \in [\SI{15}{\mu V}, \SI{167}{\mu V}$]) used to extract power-law exponent.  \textbf{(b)} Derivative of the log of $G = G - G_{\text{min}}$ divided by the derivative of the log of $V$ versus data point index. Data is fit to a constant which is plotted as the dashed blue line.  The exponent is found to be $g = 1.997 \pm 0.055$}
    \label{fig:supp_fig_power_law_fit}
\end{figure*}

It is clear from the inset of Fig.\ref{fig:fig1}C that the bias dependence appears to be well fit by parabola, with a curvature leading to a $T_0 \approx \SI{9}{K}$.  However, one would like to asses the power law exponent and its uncertainty while allowing $T_0$ to vary.  Given that we can determine the approximate range over which $G \propto V^2$ behavior would be expected, given a rough value of $T_0$, we can directly measure the power-law exponent and check for self-consistency.  Fig. \ref{fig:supp_fig_power_law_fit}A shows the same data in Fig. \ref{fig:fig1}C; the two gray circles mark the range over which the scaling exponent is extracted. The upper bound of this range is set by the curvature of the bias dependence, $T_0 \approx \SI{9}{K}$, which can be seen from Fig.\ref{fig:range_of_universality}A, at a temperature $T_{\text{probe}} = \SI{55}{mK}$, is about $\SI{170}{\mu V}$.  The lower bound, $V = \SI{15}{\mu V}$, is set by signal to noise.  To asses the quality of the power-law behavior over this range, we take the finite-difference of $\log (G - G_{\text{min}})$ and divide it by the finite-difference of $\log (V)$.  This is plotted in Fig. \ref{fig:supp_fig_power_law_fit}B versus data point index; the data is passed through a 10-point digital rolling average filter in order to reduce the noise generated by taking a numerical derivative.  If the system is well described by a power law, the value of $\mathrm{d} \log (G - G_{\text{min}}) / \mathrm{d} \log (V)$ should be a constant.  It is clear that the data in Fig. \ref{fig:supp_fig_power_law_fit}B is well approximated by a constant and the extracted fit value, $g$, over the bias range highlighted in Fig. \ref{fig:supp_fig_power_law_fit}A yields $g = 1.997 \pm 0.055$.  This indicates that fitting the data in Fig. \ref{fig:supp_fig_power_law_fit}A to a quadratic defined as $G(V) - G_{\text{min}} = \frac{1}{2}(\frac{eV}{k_b T_0})^2$ to extract $T_0$ precisely is self-consistent.

\newpage 
\subsection{Extracting the low-energy value of $T_0$ for comparison to the quantum impurity model}
The four line cuts presented in Fig. \ref{fig:strong_coupling}C represent tunneling spectra each individually associated with a $T_0$ which varies by nearly two orders of magnitude between the data sets.  Based on general renormalization group arguments, the Lagrangian given in \eqref{eq:lagrangian} becomes a better approximation to the system at low voltage biases and temperatures, relative to the energy scale set by $T_0 \propto 1/ \Gamma$ \cite{kane_transmission_1992}.  While a lower-bound on $T$ is set by cryogenic instrumentation, $V$ can be made nearly arbitrarily small.  So generally, the low-bias regime of each data set in Fig. \ref{fig:strong_coupling}C is where Eq.~\eqref{eq:lagrangian} is likely to be an accurate model of the system.  

In principle, at low voltage bias, a perturbative expansion like that in Eq.~\eqref{eq:lutt_cond} or Eq.~\eqref{eq:sommerfeld} should produce the same behavior as the full quantum impurity model in this regime within some accuracy threshold.  As a result, the $T_0$ which parameterizes the low-energy behavior also parameterizes the entire bias-dependence if Eq.~\eqref{eq:lagrangian} holds up to the strong coupling limit.  The converse of this, is that while Eq.~\eqref{eq:lagrangian} may be accurate at low voltage bias, the effect of additional operators (like electron co-tunneling) may cause deviations from the full quantum impurity model at high-bias.  This way we may assess how well the data in Fig. \ref{fig:strong_coupling}C agrees with the full quantum impurity model.   We extract $T_0$ from the low-energy (voltage bias) region of the curve, and the same $T_0$ can then be used to interpolate between the weak and strong coupling regimes via Eq.~\eqref{eq:integral_conductance}.  While these curves are all taken at a fixed electronic temperature $T_e = \SI{91}{mK}$, the large variance in $T_0$ implies that even the low-bias regime of each curve requires careful treatment in order to be an accurate low-energy expansion of Eq.~\eqref{eq:integral_conductance}.

For all curve fits, a systematic zero-bias shift of $\SI{2.5}{\mu V}$ is accounted for.  For the data sets with the largest values of $T_0$ in Fig.~\ref{fig:strong_coupling}C, i.e, $T_0 = \SI{9.02}{K}$ and $T_0 = \SI{6.87}{K}$ respectively at a base temperature of $T = \SI{56}{mK}$, there is a significant bias range where Eq.~\eqref{eq:lutt_cond} is accurate.  In this regime at low bias, $T_0$ is directly related to the curvature of the bias-dependence and is temperature independent; consequently $T_0$ can be fit agnostic of the electronic temperature for these two data sets by subtracting the residual conductance and fitting the resulting parabola (as shown in Fig.~\ref{fig:fig1}C).  As is mentioned in the main text, for these values of $T_0$, all of the temperature dependence falls in the residual conductance when $V = 0$ and we utilize this to realize a sensitive primary thermometer, using the $T_0 = \SI{9.02}{K}$, to accurately extract the electron temperature, $T_e = \SI{91}{mK}$.  This measured electronic temperature then becomes an input parameter to the remaining data sets in Fig.~\ref{fig:strong_coupling}C.  The success of this procedure can be seen in Fig.\ref{fig:strong_coupling}C where the zero-bias conductance value is well predicted for the $T_0 = \SI{6.87}{K}$ data (where $T_0$ is extracted from the curvature of $G$ agnostic of $T_e$) given the measured value of $T_e$. 

However, for the curve with $T_0 \approx \SI{1.6}{K}$ Eq.~\eqref{eq:lutt_cond} will not be accurate even when $V = 0$ when $T = \SI{91}{mK}$.  However, Fig. \ref{fig:range_of_universality}C shows that the Sommerfeld expansion in Eq.~\eqref{eq:sommerfeld}, for a temperature of $\SI{91}{mK}$, even with $T_0 \approx 1.6K$ will be within $5\%$ of the full quantum impurity model for $V < \SI{45}{\mu V}$.  To extract $T_0$, we fit this data set to the Sommerfeld expansion (in a range between $\SI{-45}{\mu V}$ and $\SI{45}{\mu V}$) given in Eq.~\eqref{eq:sommerfeld}, where $T = T_e = \SI{91}{mK}$ resulting in a fit value $T_0 = \SI{1.59 \pm 0.01}{K}$.  For the final data set in Fig. \ref{fig:strong_coupling}C, with $T_0 \approx \SI{130}{mK}$, it is clear from Fig. \ref{fig:range_of_universality}C that even the Sommerfeld expansion will not be accurate even for $V = 0$ in this regime.  The only way to systematically extract $T_0$ from this data set is to directly fit the data to Eq.~\eqref{eq:integral_conductance} where $T = T_e = \SI{91}{mK}$.  Following this procedure, $T_0 = \SI{132 \pm 2}{mK}$, and the low bias regime is well fit by this value while some deviation is observed at high-bias, further emphasizing that for low biases and temperatures, even for extremely low values of $T_0$, the system is governed by the Lagrangian in Eq.~\eqref{eq:lagrangian} to a high degree of accuracy.

\newpage 
\section{Additional Data at a different magnetic field}
We present here additional tunneling data at $B = \SI{9}{T}$ between the $\nu = 1/3$ and $\nu = 1$ edges.  All of the features presented in the main text are reproduced at a different magnetic field and set of gate voltages; this includes both the weak coupling universal scaling behavior, the strong coupling Andreev reflection of fractionalized quasiparticles, and near agreement with the full quantum impurity model. In principle, by taking Hall transport data at $V_{\text{BG}} = -2V$, we attempted to set $V_{\text{W}}$ such that the west side of the junction was set to the center of the $\nu = 1/3$ plateau.  The same procedure was taken for the data at $B = \SI{10}{T}$ in the main text.  However, uncertainty in the measured capacitance ratio between the top and bottom gates to the monolayer, would imply on general grounds that the west side of the junction sits at different locations within the $\nu = 1/3$ plateau between this data set at $B = \SI{9}{T}$ and the data presented in the main text at $B = \SI{10}{T}$. While this is not a detailed analysis of how the universal power law scaling varies versus filling between $\nu = 1/3$ and $\nu = 1/2$, where a plateau in the scaling exponent is expected, it does demonstrate the repeatability of the tunneling behavior tuning across a presumably different set of parameters including voltage bias, temperature, $T_0$, and electron density within the $\nu = 1/3$ plateau.  

\begin{figure*}[ht]
    \centering
    \includegraphics[width = 175mm]{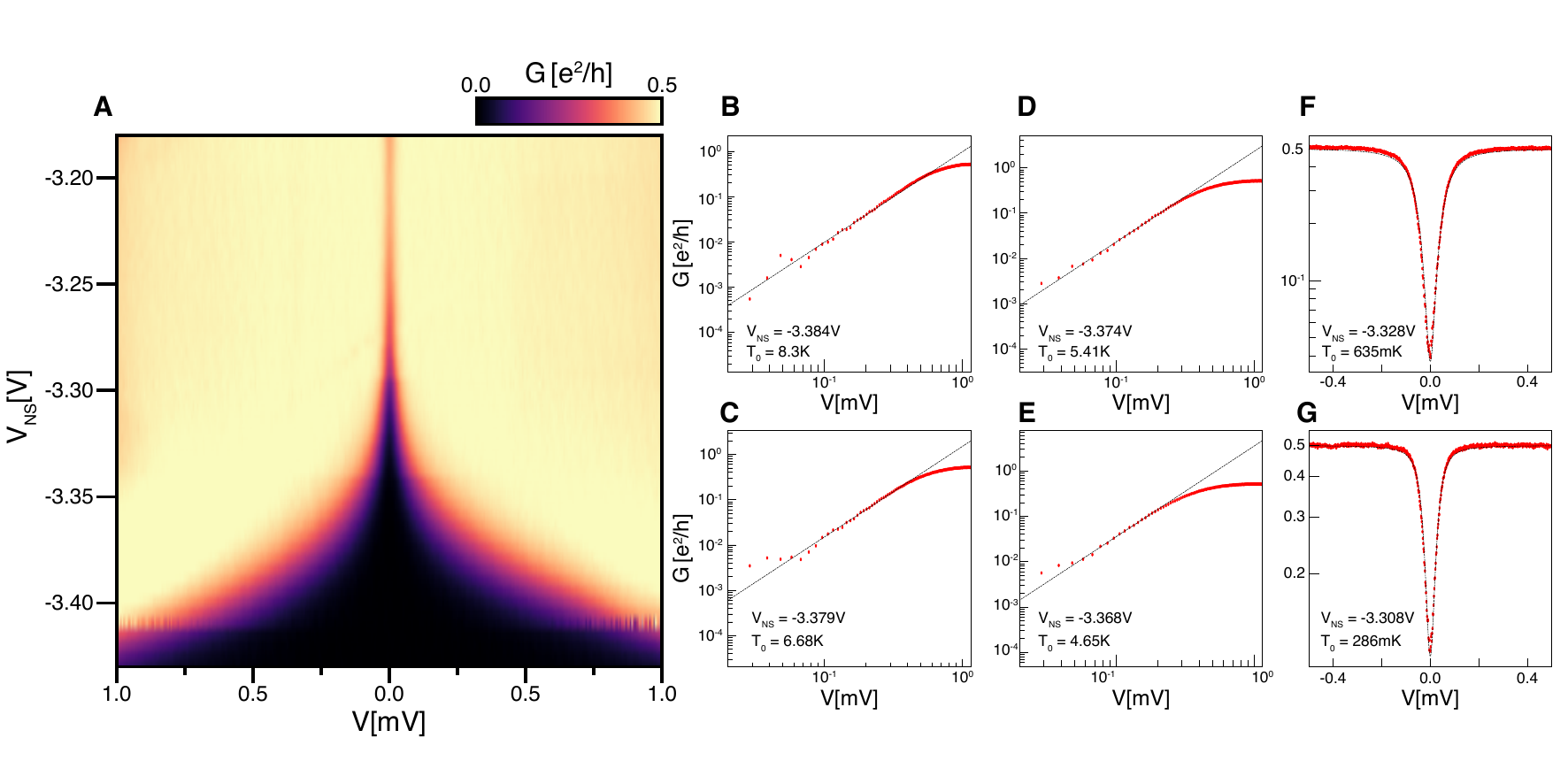}
    \caption{\textbf{Bias dependence of tunneling data between 1/3 and 1 QH edges at B = 9T.} \textbf{(a)} G plotted as a function of voltage bias $V$ and North/South gate voltage $V_{\text{NS}}$.  The other gates are held at fixed voltages: $V_{\text{BG}} = 2V$, $V_{\text{E}} = -1.46V$, $V_{\text{W}} = -1.775$.  (\textbf{B-E}) Line cuts of panel A at different values of $V_{\text{NS}}$ plotted on a log-log scale.  Data points are averaged together in windows of 160 points and a systematic zero-bias shift of $\SI{5}{\mu V}$ is removed.  Black dashed lines are fits to $G = \frac{1}{2} (\frac{e V}{k_b T_0})^2$; the fit value of $T_0$ inset within each plot.  The power-law behavior extracted from the line cuts of panel A were not taken with enough averaging to distinguish $G(V = 0)$ from 0.  Consequently, panels B-E were fit over a half-decade of bias where the lower bound was set by where the $V^2$ term is expected to dominate over the constant offset in Eq.~\eqref{eq:lutt_cond}. \textbf{(F-G)} Additional line cuts of panel A at two distinct values of $V_{\text{NS}}$ plotted on a semi-log scale.  Black dashed line is a fit to the conductivity predicted by the full quantum impurity model given in Eq.~\eqref{eq:integral_conductance}.  The fitted values of $T_0$ are inset in each plot. }
    \label{fig:9T_power_laws}
\end{figure*}

\begin{figure*}[ht]
    \centering
    \includegraphics[width = 179mm]{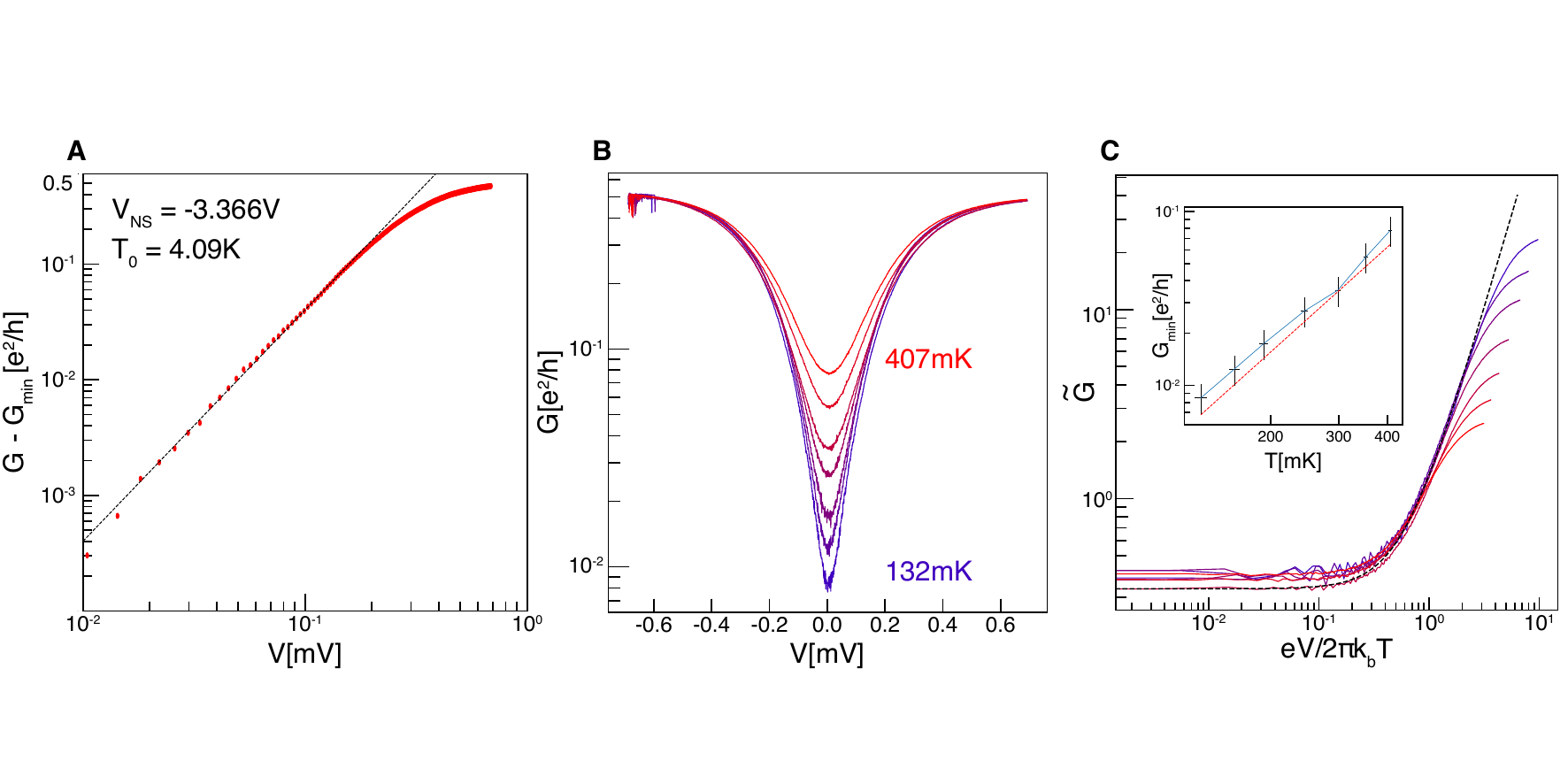}
    \caption{\textbf{Scaling collapse of tunneling data between 1/3 and 1 QH edges at B = 9T.} \textbf{(a)} $G - G_{\text{min}}$ plotted versus $V$ for $V_{\text{NS}} = -3.366V$.  The black dashed line represents the fit to the function $\frac{1}{2} (\frac{e V}{k_b T_0})^2$, yielding a fit $T_0 = \SI{4.09}{K}$.  \textbf{(b)} $G$ plotted versus $V$ for the probe temperatures: $\SI{132}{mK}$, $\SI{161}{mK}$, $\SI{192}{mK}$, $\SI{244}{mK}$, $\SI{298}{mK}$, $\SI{352}{mK}$, $\SI{407}{mK}$.  \textbf{(c)} Same curves as in panel B, but the conductance is scaled such that $\widetilde{G} = 2 G (\frac{T_0}{2\pi T_{\text{probe}}})^2 (\frac{h}{e^2})$, and is plotted against $eV/(2\pi k_b T_{\text{probe}})$ where $T_0 = \SI{4.09}{K}$ is extracted from the bias dependence shown in panel A.  Inset is $G(V = 0) \equiv G_{\text{min}}$ plotted against the probe temperature on a log-log scale.  A rough $T^2$ dependence is observed between $T = \SI{132}{mK}$ and $\SI{407}{mK}$.  The red dashed line corresponds to $G(V = 0) = \frac{1}{2} (\frac{eV}{k_b T_0})^2$, with $T_0 = \SI{4.09}{K}$. Error bars are computed assuming a current uncertainty of $\pm \SI{500}{fA}$ and a voltage uncertainty of $\pm \SI{100}{nV}$, as well as a temperature uncertainty of $\pm \SI{5}{mK}$.  While the curves collapse onto each other well, a small systematic deviation between the collapsed data set and the prediction of Eq.~\eqref{eq:lutt_cond} is observed.  This may be attributed to a systematic scaling in the zero-bias conductance as a result of insufficient averaging times near zero-bias for a fixed gate-sweep rate.}
    \label{fig:my_label}
\end{figure*}

\onecolumngrid


\end{document}